\documentclass[11pt]{article}
\usepackage[margin=1in]{geometry}
\usepackage[style=authoryear,maxbibnames=9,maxcitenames=2,backend=biber,firstinits=true,doi=false,isbn=false,url=false]{biblatex}
\usepackage{authblk}
\usepackage{color}
\usepackage{amsmath}
\usepackage[]{algorithm2e}
\usepackage{amssymb}
\usepackage{graphicx}
\usepackage{diagbox}
\usepackage{makecell}
\usepackage{caption}
\usepackage{subcaption}
\usepackage{float}

\newcommand{\bx}{\mathbf{x}}
\newcommand{\bX}{\mathbf{X}}
\newcommand{\ba}{\mathbf{a}}
\newcommand{\bb}{\mathbf{b}}
\newcommand{\bH}{\mathbf{H}}
\newcommand{\pkg}[1]{{\normalfont\fontseries{b}\selectfont #1}}
\newcommand{\mmg}{{\tt MMG}(\bH_i)}
\newcommand{\yhati}{\widehat{Y_i(0)}}
\newcommand{\yhatione}{\widehat{Y_i(1)}}
\newcommand{\tauihat}{\hat{\tau}}
\newcommand{\bepsilon}{\boldsymbol{\epsilon}}
\newcommand{\ind}{\mathbb{I}}
\newcommand{\indep}{\rotatebox[origin=c]{90}{$\models$}}
\newcommand{\bxtr}{\bx^{tr}}

\newcommand{\Ytr}{Y^{tr}}
\newcommand{\Ttr}{T^{tr}}
\newcommand{\bxts}{\bx^{ts}}
\newcommand{\xts}{x^{ts}}
\newcommand{\Yts}{Y^{ts}}
\newcommand{\Tts}{T^{ts}}
\newcommand{\bxvl}{\bx^{vl}}

\newcommand{\Yvl}{Y^{vl}}
\newcommand{\Tvl}{T^{vl}}

\newcommand{\fhat}{\hat{f}}
\newcommand{\E}{\mathbb{E}}

\newcommand{\nit}{n_{\bH_i}^t}
\newcommand{\nic}{n_{\bH_i}^c}
\newcommand{\nmmgi}{n_{\bH_i}}
\newcommand{\blambda}{\boldsymbol{\lambda}}

\let\proglang=\textsf

\title{Adaptive Hyper-box Matching for Interpretable Individualized Treatment Effect Estimation}

\author[1]{Marco Morucci\thanks{Equal contribution}} 
\author[2]{Vittorio Orlandi$^*$}
\author[3]{Sudeepa Roy}
\author[2,3,4]{Cynthia Rudin}
\author[2]{Alexander Volfovsky}
\affil[1]{Department of Political Science, Duke University}
\affil[2]{Department of Statistical Science, Duke University}
\affil[3]{Department of Computer Science, Duke University}
\affil[4]{Department of Electrical and Computer Engineering, Duke University}
\date{}
\begin{document}

\maketitle

\begin{abstract}
We propose a matching method for observational data that matches units with others in unit-specific, hyper-box-shaped regions of the covariate space. These regions are large enough that many matches are created for each unit and small enough that the treatment effect is roughly constant throughout. The regions are found as either the solution to a mixed integer program, or using a (fast) approximation algorithm. The result is an interpretable and tailored estimate of the causal effect for each unit.
\end{abstract}

\section{INTRODUCTION}
Interpretability is paramount in causal inference settings: high-stakes decisions involving medical treatments, public policies, or business strategies, are increasingly made on the basis of causal estimates from pre-existing data. Decision-makers in such settings must often be able to justify their choices for purposes of accountability, and must also be able to take advantage of all existing information in their decisions, rather than complex summaries of it -- interpretability 
plays a critical role to fulfill these needs. \textit{Matching methods} in causal inference, which match treated and control units with the same or similar covariate values, are commonly used for interpretability and mitigating bias. However, they 
can suffer from problems when human analysts manually choose the distance metric for matching: humans
are notoriously poor at manually constructing high dimensional functions. 




For matching, units with similar values of the confounding covariates should be matched together, so as to replicate the random assignment of treatment provided by a randomized experiment within each matched group \parencite{rubin1974, pearl2009}. 
Ideally, matching should be \textit{exact}, where a treated unit is matched with one or more identical control units in a matched group. However, when covariates are high-dimensional, it is 
generally impossible to find units with 
identical values of all covariates. 
Because of this, matching methods 
typically use a notion of closeness between units (e.g., a distance metric), 
that allows matches to be made approximately rather than exactly. The question then becomes how to construct 
a good distance metric. 


The choice of a distance metric for matching largely drives the interpretability and accuracy of the method. Coarsened exact matching \parencite{iacus2011,iacus2012}, for example, can require a user-defined coarsening of a high dimensional covariate space, which can be error-prone. 
Other matching methods, such as propensity score matching \parencite{rosenbaum1983} or prognostic score matching \parencite{hansen2008} are more automated in that they only require the user to select a model class, and may yield better estimates of average treatment effects. However, these techniques suffer from lack of interpretability: e.g., when one projects data onto the propensity score, the matched units may be distant from each other in covariate space, only having in common that they are equally likely to receive the treatment. Even in techniques like optimal matching  \parencite{rosenbaum1989}, the distance metric between units is an input parameter or a user-defined constraint, which is again problematic as the human analyst manually defines high dimensional distance metrics between units.

\textbf{Our Contribution~}  We propose a method for matching that provides both interpretability and accuracy without requiring humans to design the distance metric for matching. 
In particular, the approach \textit{learns an optimal adaptive coarsening} of the covariate space from a model trained on a separate training dataset, leading to accurate estimates of the treatment effect and interpretable matches. The matched group for a unit consists of all units within a \textit{learned} unit-specific high dimensional hyper-box.
These hyper-boxes
are constructed so that they 1$.$ contain enough units for reliable treatment effect estimates, and also so that 2$.$ 
units within each box have similar potential outcomes, which lowers the bias of the estimated treatment effect. 
This allows us to avoid black-box summaries (propensity or prognostic scores) and ad-hoc pre-specified metrics
given by the users.  
Our estimates are interpretable. First, they are case-based: each individual's estimate can be explained in terms of the units they are matched with. Second, the choice of cases is itself interpretable: if two units are matched together, it is because they fall in the same easily-described hyper-box.

We formulate the problem of learning optimal partitions for matching as an optimization problem, to which we propose two solutions. Broadly, the optimization problem solves the following minimization:
$$\min_\textrm{box}\left[\begin{array}{l}\textrm{variability}(\textrm{predictions in box}) +\\\quad 
\textrm{error}(\textrm{estimates of counterfactuals within box})\end{array}\right]
$$
subject to the constraint that the box contains at least $m$ control units when estimating causal effects for a single treatment unit (the choice of $m$ depends on the application).

\begin{figure}   
    \centering
    \includegraphics[width=0.5\columnwidth]{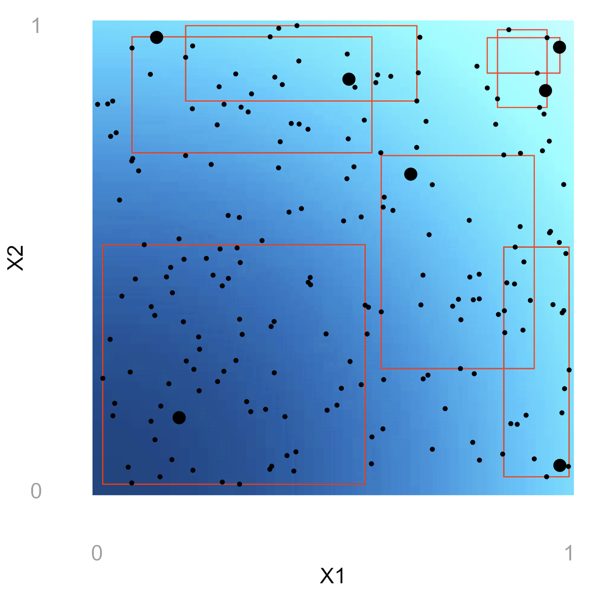}
    \caption{A toy two-dimensional dataset with covariates $X1, X2$, with a few of the matched groups shown as boxes. Each unit has its own matched group, which can overlap with others. The background indicates the true outcome values, with darker regions representing lower outcomes. The boxes are small where outcomes change rapidly, and large in regions of near-constant outcome.}
    \label{fig:toy}
\end{figure}

By training hyper-boxes in a way that leverages a model trained on a training set, we 
are able to create boxes that adapt
to the covariate space. There is a tradeoff in the construction of the hyper-boxes between including a large number of points within the hyper-box and keeping variance low for the predictions within the hyper-box; both goals can help preserve the quality of treatment effect estimation. As a result of these goals, hyper-boxes can be arbitrarily large along covariates that are {\em irrelevant} for treatment effect estimation, whereas box-widths can be small in regions where the outcome changes rapidly. Figure \ref{fig:toy} shows an example of these adaptively-learned hyper-boxes for a two-dimensional dataset. By looking at the shapes of these boxes, one can observe where the outcome changes rapidly (regions with the smaller boxes) and where it changes slowly (regions with larger boxes).

We provide two optimization methods for the boxes. First, we formulate the problem as a {\em mixed integer program} (MIP) and are thereby able to solve it exactly using 
state-of-the-art MIP solvers, which are fairly efficient for this problem. Second, we propose a faster and more scalable approximation algorithm.

In Section \ref{sec:methodology}, we present motivation, discuss issues with existing approaches to coarsening, formulate our method as a MIP, and introduce a fast approximation. In Section \ref{sec:experiments}, we compare to other matching methods in a simulation study. In Section \ref{sec:application}, we apply our method to a study of the effect of a work training program on future earnings. We conclude with a discussion in Section \ref{sec:discussion}. Our method is called ``{\em Adaptive Hyper-Box}'' ({\em AHB}) matching.

\subsection{RELATED WORK}
There is a large literature on estimating treatment effects in observational studies \parencite{stuart2010matching}, and in particular, on matching methods
\parencite[e.g.,][]{zubizarreta2012using,pimentel2018optimal,keelePimentel,de2016evaluation,rosenbaum2017imposing}. 

One formulation of our approach relies on solving a mixed integer program (MIP). MIPs have previously been used for causal inference in order to accomodate linear balance constraints on the covariates \parencite{zubizarreta2012using,zubizarreta2014matching,morucci2018hypothesis}. Our goals are entirely different from those of other MIP-based causal problems. 

There are also machine learning methods for estimating treatment effects with continuous confounders  \parencite[e.g., double machine learning,][]{chernozhukov2017}, that are not interpretable. The black box methods with the best current performance have been demonstrated to be variants of Bayesian Additive Regression Trees (BART) \parencite{hill2011bayesian,hahn2020bayesian,hill2020bayesian}.
Our method leverages a black box machine learning model (in our case, a BART model) loosely to help define hyper-boxes, using the help of the training set. 

Our work is closely related to several threads in the literature: 1. prognostic scores \parencite{hansen2008,stuart2013prognostic}, as we leverage predictions to create matches; 2.  methods within the almost-exact-matching (AEM) framework (FLAME, DAME, and MALTS) \parencite{FLAME,DAME,malts} that leverage a training set for matching, and 3. the causal forest (CF) framework \parencite{wager2018}, because they use a training set for assisting with ``soft'' matching on a test set. Matching on the prognostic score attempts to find a low dimensional summary to match on, which our approach avoids. Our method differs from FLAME and DAME (which handle only discrete covariates and use learned Hamming distances), differs from MALTS (which uses learned Mahalanobis distances on continuous covariates), and differs from CF (because it aims to specifically generate interpretable matched groups). Adaptive Hyper-Boxes handles both continuous and discrete variables in the same framework, and needs only to pinpoint hyper-box edges. 
We do not use nearest neighbors, we do not parameterize a distance metric; we use all points within the learned interpretable hyper-box.

Hyperboxes have been used extensively for regression \parencite[e.g.,][]{peters2011granular}, classification \parencite[e.g.,][]{XuPapa09} and prediction \parencite[e.g.,][]{boxclassification} but notably, not for causal inference \parencite{khuat2019hyperbox}. These methods  \parencite[and others, such as bump hunting, ][]{FriedmanFi99}
aim to find adaptive boxes around individual units
and some use MIPs to find boxes, as we do. Some other works aim to create global rule-based classifiers for causal inference \parencite{WangRu18}, whereas our method provides local rules.

\section{METHODOLOGY} \label{sec:methodology}
Throughout, we consider $n$ units and $p$ covariates. The units  are indexed by $i=1,\dots, n$,
and the covariates of unit $i$ are denoted by a $p$-dimensional random variable $\bX_i$, taking values $\bx_i = (x_{i1}, x_{i2}, \dots, x_{ip})’ \in \mathbb{R}^p$. A unit’s potential outcomes are given by $(Y_i(0), Y_i(1))$, which are also random variables in our setting. We use the following model for the potential outcomes: $Y_i(t) = f_t(\bX_i) + \nu_i$, where $\E[\nu_i]=0$, and, for any two units $i$ and $k$, $\nu_i$ and $\nu_k$ are independent. We require $f$ to be nonparametrically estimable from the data. We denote treatment by the random variable $T_i \in \{0, 1\}$; we refer to units with $T_i=1$ as treated units, and to units with $T_i=0$ as control units. We denote observed outcomes with the random variable $Y_i = Y_i(1)T_i + Y_i(0)(1-T_i)$. Our quantity of interest is the Individual Treatment Effect (ITE) for each treated unit, defined as $\tau_i = \E[Y_i(1) - Y_i(0)|\bX_i=\bx_i]$. By definition of $Y_i$, we never have access to $Y_i(0)$ for treated units, and control units must be employed to construct an estimate of this missing potential outcome for treated units.
To do this we make the following canonical assumptions of observational inference:\\
\textbf{(A1) Overlap}. For all values of $\bx$ and units $i$, we have $0 < \Pr(T_i = 1 | \bX_i = \bx) < 1$. \\
\textbf{(A2) SUTVA}. A unit's potential outcomes depend only on the treatment administered to that unit, i.e., if $Y_i(t_1, \dots, t_n)$ denotes unit $i$'s potential outcome as a function of all $n$ units' treatment status, under SUTVA we have: $Y_i(t_1, \dots, t_n) = Y_i(t_i)$.\\
\textbf{(A3) Conditional ignorability}. For all units $i$ and any $t \in \{0,1\}$, treatment is administered independently of outcomes conditionally on the observed covariates, i.e.,  $T_i \indep (Y_i(1), Y_i(0))|\bX_i=\mathbf{x}_i$. This directly implies that $\E[Y_i|T=t, \bX_i=\bx_i] = \E[Y_i(t)|\bX_i=\bx_i]$, which enables us to estimate treatment effects on observed data.
\par
Under these assumptions, if for a treated unit $i$ there existed a control unit $k$ such that $\bx_i = \bx_k$, then we would have $\E[Y_i(0)|\bX=\bx_i] = f_0(\bx_i) = f_0(\bx_k) = \E[Y_k(0)|\bX=\bx_k]$, and the estimator $Y_i - Y_k$ would be unbiased for $\tau_i$. Unfortunately, this is almost never the case in practice: since $\bx$ is high-dimensional, it is unlikely that most units would have a match with the same exact covariate values. To remedy this issue, we match treatment units to control units with similar values of $\bx$. 

\subsection{PRINCIPLES OF APPROXIMATE MATCHING VIA HYPER-BOXES}
We focus without loss of generality on creating hyper-boxes for treatment units; any control unit within treatment unit $i$'s box will be considered to be matched to $i$. Each hyper-box is $p$-dimensional. Hyper-boxes for control units can be constructed analogously.

Hyper-boxes are specified by lower and upper bounds for all covariates $\ba_i = (a_{i1}, a_{i2}, \dots, a_{ip})’$ and $\bb_i = (b_{i1}, b_{i2}, \dots, b_{ip})’$. For convenience, we define the function $H(\ba, \bb) = [a_{1}, b_{1}] \times \dots \times [a_{p}, b_{p}]$ and also denote unit $i$'s $p$-dimensional hyper-box as $\bH_i = H(\ba_i, \bb_i)$. Necessarily, $\bx_i \in \bH_i$; i.e., unit $i$ is contained in its own box. Similarly, we say that a unit $k$ is contained in $i$'s box if $\bx_k \in \bH_i$ and we define the \textit{main matched group} for treated unit $i$ to be the set of all units contained in $i$'s box: $\mmg = \{k \in 1, \dots, n: \bx_k \in \bH_i\}$. We also use $\nit = \sum_{k \in \mmg} T_k$ and $\nic = \sum_{k \in \mmg} 1-T_k$ to denote the number of treated and control units in unit $i$'s box respectively, as well as $\nmmgi = \nit + \nic$.

We use the following estimators for outcomes of unit $i$. We emphasize that both quantities are estimated from a single box associated with unit $i$; the first from control units and the second from treatment units.
\begin{align}
    \yhati &= \frac{1}{\nic}\sum_{k \in \mmg}Y_k(1-T_k). \label{Eq:CFest}\\
    \yhatione &= \frac{1}{\nit}\sum_{k \in \mmg}Y_k(T_k). \label{Eq:CFest1}
\end{align}
There are then two options to estimate $\tau_i$: $\tauihat_a = \yhatione - \yhati$, and $\tauihat_b = Y_i(1) - \yhati$. The first option is better when wanting to extend the estimated effects to a super-population of interest, as it can lower the population variance of the estimated response function, while the second option is better in finite-sample inference settings. 
It is clear by definition of our quantity of interest, $\tau_i$, that our objective should be constructing hyper-boxes for unit $i$ such that $\yhati \approx Y_i(0)$, and $\yhatione \approx Y_i(1)$. 

We thus follow three principles in creating hyper-boxes:\\
    1. \textit{Bias Minimization:} Matches should yield high quality estimates of the treatment effect. To this end, we create large boxes with low variance in their estimates. 
    2. \textit{Interpretability:} Matches must be interpretable to permit case-based reasoning. 
    3. \textit{Honesty:} No test outcomes  may  be  used to construct hyper-boxes. This helps lower bias, and is a general principle of causal inference \parencite{rubin2005, wager2018}. We may use covariates and outcomes of a separate training set, and covariates for the (test) units to be matched.
\paragraph{Issues with existing fixed-width coarsening methods.}
Common matching methods based on pre-specified fixed-width bins \parencite{iacus2011, iacus2012}, will take as input a desired box size for each covariate, $\bepsilon = (\epsilon_1, \dots, \epsilon_p)$, and then construct boxes of size exactly $\|\bepsilon\|_1$. This approach suffers from two issues:\\
\textbf{Issue 1:} $\|\bx_i - \bx_k\|_1 \geq \|\bepsilon\|_1$, but $|f_t(\bx_i) - f_t(\bx_k)|$ is small. In this case we have two units that are further away on the space of $\bx$ than the pre-specified tolerance, but it is entirely possible that these units could have similar values of the outcome function.
In this case, the units would not be matched, leading to few (or no) matches for $i$ and therefore a poor (or nonexistent) ITE estimate. \\
\textbf{Issue 2:} $\|\bx_i - \bx_k\|_1 \leq \|\bepsilon\|_1$, but $|f_t(\bx_i) - f_t(\bx_k)|$ is large. This could happen in the case in which $\bepsilon$ is pre-specified without taking variation in the response function into account. If the slope of the response function is large, then even units that have close values of $\bx$ will have significantly different values of $y(0)$. Matching $i$ to $k$ in this case would lead to a bad estimate of $i$'s ITE.

\begin{figure}[t] 
    \centering
    \includegraphics[width=\columnwidth]{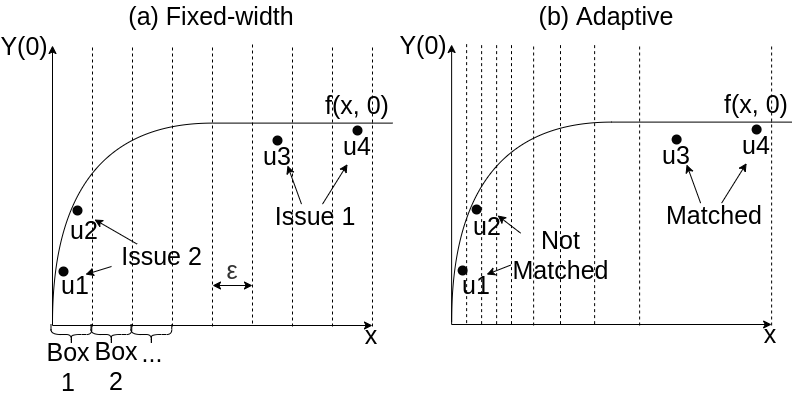}
    \caption{Issues from matching with fixed-width boxes are demonstrated in Panel a. The solid line represents the outcome function, black dots are units to be matched, and vertical dashed lines represent fixed-width boxes. Issue 1 arises when u3 and u4 are not matched together because they are in different boxes, despite having almost constant values of $Y$ within the full range between them. Issue 2 is present because  u1 and u2, matched together (as they are in the same box), have different values of $Y$. These issues are absent when boxes are made adaptively to the outcome function, as demonstrated in Panel b.
     \label{fig:issues}}
\end{figure}

Several rules have been developed to choose fixed-width bins based on the data \parencite[e.g.,][]{scott1979, freedman1981, wand1997}. These rules do not take into account relationships between covariates and outcome, and are thus vulnerable to the two issues above.

\subsection{THE ADAPTIVE HYPER-BOX FRAMEWORK}
Our proposed framework aims at creating bins for interpretable adaptive matching, avoiding the issues discussed above. Instead of starting from a pre-specified value of box size, $\epsilon$, we learn unit-specific boxes from the data itself, by directly minimizing quantities related to the principles outlined previously. We aim for hyper-boxes that solve the following optimization problem:
\begin{align*}
    \min_{\bH_1,\dots,\bH_n}& \sum_{i=1}^n \operatorname{\operatorname{Err}}(\bH_i) + \operatorname{Var}(\bH_i)\\
    \mbox{Subject to: }& \nmmgi \geq m\; \forall\, i,  
\end{align*}
where Err and Var are as in Eqs.~\eqref{eq:err}-\eqref{eq:var1}. In words, we would like to minimize bias and variability of each box, while making sure that at least $m$ units are contained in each hyper-box. To minimize bias, we would like boxes that contain units whose observed outcomes are strongly predictive of the missing control outcome of interest. This can be achieved by defining error as follows:
\begin{align}\label{eq:err}
    \operatorname{Err}(\bH_i) &= \biggl| f_0(\bx_i) - \frac{1}{\nmmgi}\sum_{k \in \mmg} f_0(\bx_k)\biggr|\nonumber\\
    &+ \biggl| f_1(\bx_i) - \frac{1}{\nmmgi}\sum_{k \in \mmg} f_1(\bx_k)\biggr|.
\end{align}
For reliable estimates, we encourage boxes to contain (1) a large number of units, and (2) to minimize variability of predicted outcomes on the control units it contains: 
\begin{align}\label{eq:var1}
    &\operatorname{Var}(\bH_i) = \nonumber\\
    &\frac{1}{\nmmgi}\sum_{k \in \mmg}\left(f_0(\bx_k)  - \frac{1}{\nmmgi}\sum_{k \in \mmg} f_0(\bx_k)\right)^2 \nonumber\\
    +& \frac{1}{\nmmgi}\sum_{k \in \mmg}\left(f_1(\bx_k)  - \frac{1}{\nmmgi}\sum_{k \in \mmg} f_1(\bx_k)\right)^2.
\end{align}
Minimizing $\operatorname{Err}(\bH)$ and $\operatorname{Var}(\bH)$ directly avoids Issues 1 and 2 outlined above. In the case of Issue 1, both $\operatorname{Err}(\bH)$ and $\operatorname{Var}(\bH)$ will be small even if units are far apart in terms of $\bx$, telling us that we can make boxes larger in that part of the space. In the case of Issue 2 the opposite will be true; even if units are close in terms of $\bx$, $\operatorname{Err}(\bH)$ and $\operatorname{Var}(\bH)$ will be large, suggesting that boxes should be smaller in that part of the space. 

Our loss will be reliable if we have good estimates $f_t(\bx)$ at many points within each bin, including all points $\bx_1, \dots, \bx_n$ at a minimum. We preserve honesty in such estimates by dividing the data into a training and a test set, denoted by $D^{tr} = \{(\bxtr_i, \Ytr_i, \Ttr_i)\}_{i=1}^n$ and  $D^{ts} = \{(\bxts_i, \Yts_i, \Tts_i)\}_{i=1}^n$ respectively, and assumed to each be of size $n$ for notational simplicity. Lastly, under these conditions, the hyper-boxes are designed to provide balance on relevant covariates and thus lead to high quality treatment effect estimates \parencite{stuart2013prognostic}. The test set will contain the observations to be matched, while the training set will be used to estimate $f_t(\bx)$ for each $\bx$ of interest. We will denote this estimate by $\fhat_t(\bx)$: any machine learning method can be used to estimate $f_t$, as predicted values of $f_t$ are only going to inform loss calculations and not actual treatment effect estimates. 

%
%
\paragraph{Adaptive Hyper-box MIP formulation.} 
We use the triangle inequality to upper-bound the error term as follows. Here, we consider treatment point $i$, and points $k$ within its main matched group $\mmg$, for an arbitrary treatment value, $t$, and hyper-box $\bH_i$: 
\begin{align}
    Err(\bH_i) &= \biggl|\frac{1}{\nmmgi}\sum_{k \in \mmg}f_t(\bx_i) - f_t(\bx_k)\biggr|\nonumber\\
    &\leq \frac{1}{\nmmgi}\sum_{k \in \mmg}\biggl|f_t(\bx_i) - f_t(\bx_k)\biggr|\label{Eq:lossbound}.
\end{align}
We minimize the bound instead of the error term, for both treatment and control groups. We use a similar upper bound for variability. For any value of $\bH_i$ we have:
\begin{align*}
    &Var(\bH_i)
    \\&= \frac{1}{\nmmgi}\sum_{k \in \mmg}\left(f_t(\bx_k) - \frac{1}{\nmmgi}\sum_{l \in \mmg}f_t(\bx_l)\right)^2 
    \\ &\leq \frac{1}{\nmmgi}\sum_{k \in \mmg}C\left|\frac{1}{\nmmgi}\sum_{l \in \mmg}(f_t(\bx_k) - f_t(\bx_l))\right|,
\end{align*}
where the last line follows by setting $C=\max_{\bH_i}\left|\frac{1}{\nmmgi}\sum_{l \in \mmg}(f_t(\bx_k) - f_t(\bx_l))\right|$ and using H\"older's Inequality. Here $C$ is a constant, and is not affected by any optimization we will perform to obtain $\bH_i$. We can now apply the triangle inequality twice: 
\begin{align}
    &Var(\bH_i) \nonumber
    \\ & \leq \frac{1}{\nmmgi}\sum_{k \in \mmg}\frac{C}{\nmmgi}\sum_{l \in \mmg}|f_t(\bx_k) - f_t(\bx_i)| \nonumber
    \\ & \mspace{214mu} + |f_t(\bx_l) - f_t(\bx_i)| \nonumber
    \\ &= \frac{2C}{\nmmgi}\sum_{k \in \mmg}|f_t(\bx_k) - f_t(\bx_i)|.\label{Eq:varbound}
\end{align}
Looking at \eqref{Eq:lossbound} and \eqref{Eq:varbound}, we see that minimizing $\sum_{k \in \mmg}|f_t(\bx_i) - f_t(\bx_k)|$ will lower both $Err(\bH_i)$ and $Var(\bH_i)$ through the upper bounds just introduced, for fixed $n_{\bH_i}$. Minimizing this term also ensures that the treatment and control outcomes both stay relatively constant within each learned hyper-box. 

In order to ensure that the denominator of the variance (i.e., $n_{\bH_i}$) stays large, we subtract it from the loss function. Hence, the loss now encourages larger matched groups, while maintaining linearity of the objective:
\[
\min_{\bH_i} \sum_{k \in \mmg}|f_t(\bx_k) - f_t(\bx_i)| + \beta\nmmgi,
\]
where $\beta$ trades off between the terms.

These steps give rise to the following global MIP for our entire sample. Here, decision variable $\bH_i$ defines the box for treatment unit $i$, and decision variable $w_{ik}$ is an indicator for whether $k$ is in $i$'s box:
\begin{eqnarray}\nonumber
\lefteqn{   \hspace*{-2pt}\min\limits_{\bH_1,\dots,\bH_n}\sum_{i=1}^n\biggl\{
    \gamma_1\sum_{k=1}^nw_{ik}\left|\fhat_1(\bxts_i) - \fhat_1(\bxts_k)\right|\nonumber
    }\label{Eq:MIPloss}\\
    \hspace*{-4pt}&&\hspace*{-4pt}+\gamma_0\sum_{k=1}^nw_{ik}\left|\fhat_0(\bxts_i) - \fhat_0(\bxts_k)\right|-\beta\sum_{k=1}^n w_{ik}\biggr\}
    \end{eqnarray}
    \begin{align}
    \mbox{subject to:}&\; \bH_i \in \mathbb{R}^{p\times p}, w_{ik} \in \{0,1\}\; &&\forall\, k \nonumber\\
    & \xts_i \in \bH_i &&\forall\, i\label{Eq:MIPC1}\\
    & w_{ik} = \ind_{[\xts_k \in \bH_i]}&&\forall \, i\label{Eq:MIPC2}\\
    & \sum_{k=1}^n w_{ik}(1-T_k) \geq m&&\forall\, i.\label{Eq:MIPC3}
\end{align}
Constraint \eqref{Eq:MIPC1} forces unit $i$ to be within its own box; \eqref{Eq:MIPC2} defines an indicator $w_{ik}$ for whether unit $k$ falls into the box for test unit $i$; \eqref{Eq:MIPC3} forces boxes to include at least $m$ control units. We require a minimum number of control, but not treatment, units to be matched, because treatment unit $i$ is within $\mmg$, and thus there is always at least one treated unit in each box. This makes computing the first term in the loss always possible, and excludes trivial solutions with empty boxes. The loss in Eq. \eqref{Eq:MIPloss} is made up of three terms: the first is the upper bound on the estimation error and variability terms of our framework derived in inequalities \eqref{Eq:lossbound} and \eqref{Eq:varbound} for treated outcomes. The second is the same bound, but for control outcomes. We want these terms to be small to ensure the outcome function does not vary much within a box. The third term counts units in the box, encouraging more matches. 
The supplement details an explicitly linear formulation of the above problem. The hyperparameters $\gamma_1$, $\gamma_0$, and $\beta$ weight the three components of the loss. They can be cross-validated, set to 1, or chosen intuitively by normalizing them to the same scale as discussed in the supplement. 

The form of the MIP presented above directly suggests that the optimization problem is separable in the $1\dots, n$ units. We take advantage of this property and solve one MIP for each of the $n$ units to be matched separately. 
\paragraph{Adaptive Hyper-box Fast Approximation}
We now describe a fast algorithm to approximate the MIP solution. For a unit $i$, we initialize its box to be a single point at its covariate values. We then expand the box according to the principles previously outlined: 1. we expand the box along a single covariate at a time, so that the resulting box is always axis-aligned and interpretable; 2. we expand along the covariate that extends the box into the region with least outcome variation -- ensuring high quality matches -- and stop expanding the box once this variation increases too much, avoiding low quality matches; and 3. we estimate the variation in the outcome via $\hat{f}_0, \hat{f}_1$ learned on a separate, training set, as for the MIP. 

Algorithm \ref{greedy_algo} in the supplement provides pseudocode. The main crux of the algorithm is to determine whether a new, proposed box $\mathbf{P}$ is good. To do so, we examine the outcome function in $\mathbf{P}\backslash\bH_i$ (the region we propose to add to our existing hyper-box). If the outcome in the new region is relatively constant, we do not expect to incur much bias from including units that lie inside. Therefore, we look at how much $\hat{f}_0, \fhat_1$ vary on a grid in $\mathbf{P}\backslash\bH_i$ and choose to expand along the covariate yielding the lowest variation. Further details are in the supplement.

\paragraph{Scalability and Parallelization}
Both MIP AHB and Fast AHB create a box tailored to a specific unit $i$, independently from boxes of other units. Both methods are, therefore, embarassingly parallelizable. The supplement shows runtime results for the methods: Fast AHB scales well, especially in $n$, and can be applied to large datasets on most machines, while MIP AHB is less suited for large datasets due to its exponential nature. Discussion of the methods' computational complexity, and suggestions for speeding them up, is included in the supplement. 

\paragraph{Matching with Non-Continuous Covariates} 
Our method also handles non-continuous covariates, including categorical and cardinal covariates. Categorical covariates that take on $k$ discrete values can be binarized into $k - 1$ indicator variables, after which MIP and Fast can be run without modification to form matches. MIP and Fast can also be run out of the box on  cardinal variables without loss in performance. We demonstrate this by matching on year-valued variables in our application. 

Empirically, when we run MIP AHB and Fast AHB on categorical data, they learn identical importance weights for the covariates (see Section \ref{sec:exp2}). That is, they either construct boxes that exactly match units with identical covariate values or prioritize matches on covariates contributing more to the outcome. This is similar to the characteristics of the FLAME and DAME algorithms described by \cite{FLAME} and \cite{DAME}, though AHB has the added benefit of adaptively handling continuous covariates. It would not be possible to extend FLAME and DAME to this case because they rely on Hamming distance. Since AHB chooses only box edges, it avoids having to use a parameterized distance metric, allowing it to handle continuous covariates in the \textit{same way} that it handles discrete covariates.

\section{EXPERIMENTS}\label{sec:experiments}
We generate data independently for all units, with data for unit $i$ generated according to the following process: 
\\1. Generate covariates: $x_{ij} \stackrel{ind}{\sim} F_x, j = 1, \dots, p$\\2.  Generate a propensity score: $e_i = \textrm{expit}(\boldsymbol{\gamma}\bx_i)$\\3. Assign treatment: $Z_i \sim \operatorname{Bernoulli}(e_i)$\\ 4. Generate the outcome: $y_i = g(\bx_i) + h(\bx_i) Z_i + \epsilon_i$.\\

Here, $\boldsymbol{\gamma}$ is fixed. We consider various choices of confounding functions $g$ and heterogeneous treatment functions $h$, seen in Table \ref{tab:sim_setup}, subject to which we evaluate estimation of the ITE of treated units. All results are averages across 10 simulations, each with $n = 600$ units. The supplement contains additional simulations studying higher dimensional settings, correlated covariates, and coverage of ITE confidence intervals.

We compare the following estimators:
\textbf{BART} - Bayesian Additive Regression Trees \parencite{bart,hill2011bayesian} estimates $\text{ITE}_i$ as $\hat{f}_1(\bx_i) - \hat{f}_0(\bx_i)$, 
\textbf{Best CF} - $1:k$ matching of a treated unit $i$ to the $k$ control units with outcomes closest to $i$'s true counterfactual (this is cheating: one does not have this extra information in practice), 
\textbf{GenMatch} - Genetic Matching of a treated unit to at most $k$ control units \parencite{genmatch}, 
\textbf{CEM} - Coarsened exact matching \parencite{iacus2011, iacus2012}, 
\textbf{Propensity Matching} - $1:k$ propensity score matching, 
\textbf{Prognostic Matching} -  $1:k$ prognostic score matching, 
\textbf{Full Matching} - Full matching \parencite{fullmatch}, 
\textbf{Mahal} - $1:k$ matching on the Mahalanobis distance between covariates, 
\textbf{Fast} - Our proposed approximate algorithm for AHB, 
\textbf{MIP} - Our proposed MIP for AHB. 

For all $1:k$ matching estimators, we consider $k \in \{1, 3, 5, 7, 10\}$ and report the best results attained. All nearest neighbor matching is performed with replacement. BART, Prognostic Matching, Fast, and MIP first split the data and fit BART on the training set to estimate an outcome model. For AHB, boxes are then constructed from the outcome model to be used on the test units. In addition to using BART to power Prognostic, MIP, and Fast, we include the BART estimator to directly predict counterfactuals for units in the test set. In this way, we compare our approach to the limits of predictive performance attainable using a highly flexible -- and highly uninterpretable -- method. Similarly, we include the Best CF estimator to compare to performance attainable when using counterfactual data that is unobserved in practice. We defer details of implementations to the supplement.

\subsection{
CONTINUOUS COVARIATES}
First, we assess our method's performance in settings where different functions of continuous covariates confound the outcome and modulate the treatment effect. We simulate $x_{ij} \stackrel{ind}{\sim} U(0, 1)$ and choose $g$ (confounding function) and $h$ (heterogeneity function) as specified in the first six rows of Table~\ref{tab:exp_results}. Below, we label simulation settings as ``Confounding function / Treatment function''. MIP or Fast perform better than all other methods in all but the None / Const and Linear / Const setups, where BART outperforms us. This is reasonable given its highly flexible (yet uninterpretable) nature. 

MIP and Fast perform well even when there is a heterogeneous treatment effect in addition to confounding (row 6 of Table \ref{tab:exp_results}). Actually, MIP and Fast tend to outperform competing ones by greater margins when heterogeneous treatment is introduced on top of confounding, as can be seen by comparing the Box / Const and Box / Box setups. 

\begin{figure}
    \centering
    \includegraphics[width=0.5\columnwidth]
    {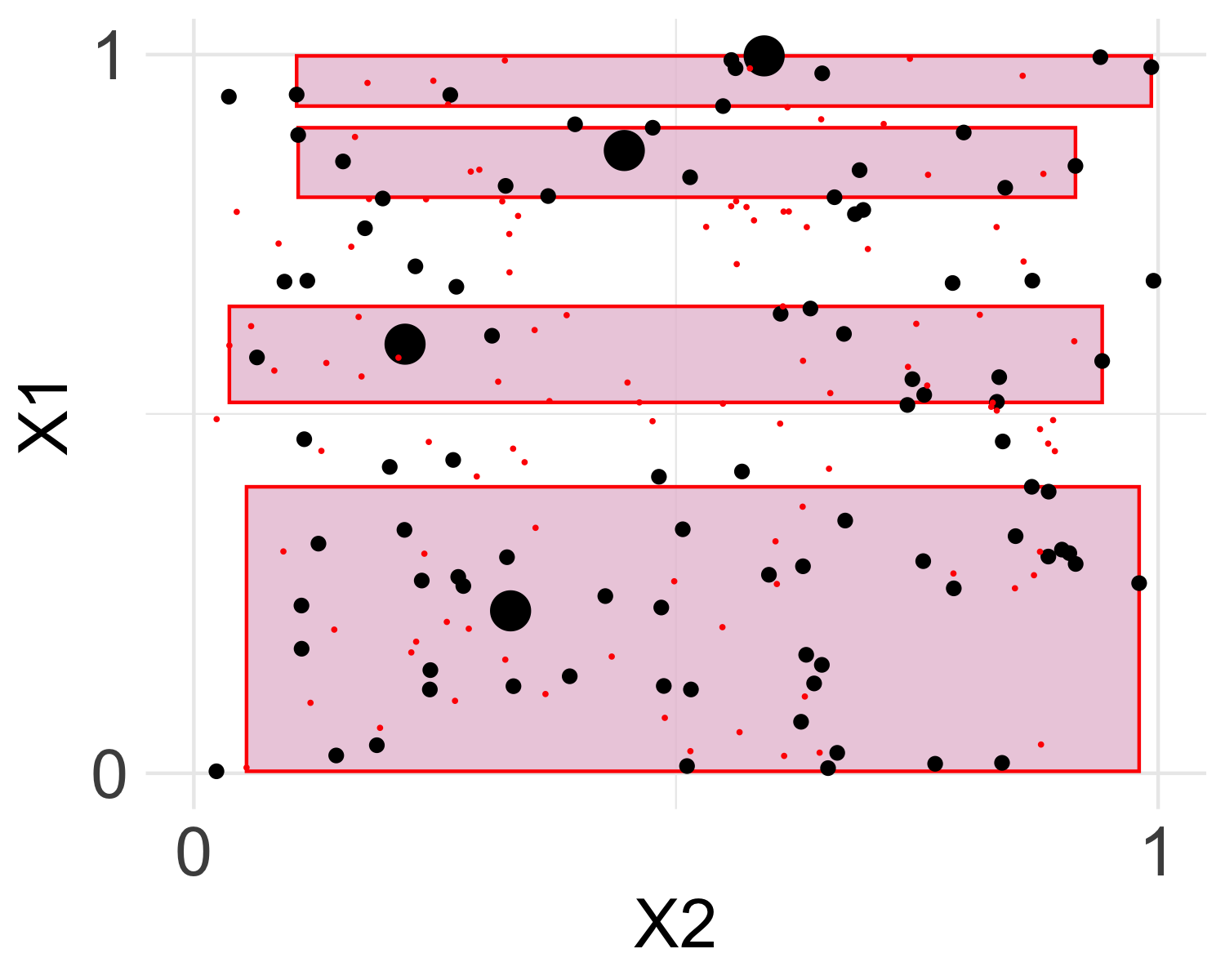}
    \caption{The boxes formed by Adaptive Hyper-Boxes for four example points (enlarged). The box widths span most of the horizontal axis, associated with an irrelevant covariate. The height of the boxes decreases moving upwards, as confounding increases. Black and red points denote treatment and control units, respectively.}
    \label{fig:bin_viz}
\end{figure}

When there are irrelevant covariates (e.g. row 5 of Table \ref{tab:exp_results}), CEM fails to make even a single match due to the high dimensionality of the space. On the other hand, AHB adapts to the irrelevant covariates; we can visualize this by examining in Figure \ref{fig:bin_viz} some of the boxes it learns for setup Quad / Const. The vertical axis represents the one covariate relevant to the outcome and the horizontal axis an arbitrary irrelevant covariate. We see that AHB learns which covariate is important: it makes the boxes skinny along one dimension -- as they should be sensitive to the changes in outcome along that axis -- and expand fully throughout the range of the other irrelevant dimension. The height of the boxes also decreases along the vertical axis, because the effect of confounding on unit $i$ is given by $x_{i1} ^ 2$. Variation in $x_{i1}$ therefore has greater impact on the outcome near 1 than near 0 and the boxes reflect this.

\subsection{
DISCRETE AND MIXED COVARIATES}\label{sec:exp2}
Here, we evaluate the performance of our method on discrete and mixed (discrete and continuous) data. Abusing notation slightly, we will use $x$ to refer to continuous covariates, of which there will be $p_c$, and we use $w$ to refer to discrete covariates, of which there will be $p_d$. We consider binary covariates, because we can binarize any $k$-level discrete covariate into $k - 1$ indicator variables, allowing us to match on any subset of the $k$ levels. Binary covariates are simulated $w_{ij} \stackrel{ind}{\sim} \text{Bernoulli(0.5)}$. 

Choices of $g$ and $h$ and associated results are specified in the last three rows of Table \ref{tab:exp_results}. We see that CEM and AHB both perform exceedingly well when all covariates are binary. Further analysis reveals: 1. that MIP and Fast yield identical boxes and ITEs in this scenario, 2. that the ITEs are the same as those generated via exact matching \emph{on the one, true covariate}, and 3. that CEM's ITEs are the same as those generated via exact matching \emph{on all covariates}. Thus, while both methods yield unbiased ITE estimates in this setting, CEM's are of higher variance because it constructs more granular boxes than necessary due to its inability to adapt to irrelevant covariates. Indeed, supplemental results show that as the number of irrelevant covariates increases, CEM's performance deteriorates drastically, while AHB's stay the same. In the simulation with mixed covariates, MIP AHB outperforms all competitors but BART, and Fast AHB falls only behind BART, Best CF, and Prognostic. 

\paragraph{Similarity Between MIP AHB and Fast AHB}
To compare MIP AHB and Fast AHB, we compare the overlap in units assigned to matched groups by MIP AHB and Fast AHB, denoted by $\mmg^{\text{MIP}}$ and  $\mmg^{\text{Fast}}$. We define a `mutual membership rate' as the maximum of the proportion of units in $\mmg^{\text{MIP}}$ that are in $\mmg^{\text{Fast}}$ and vice versa. Across all units, we find median mutual membership rates around 80\% in our experiments. Visual comparisons of the boxes output by both methods also confirm they adapt similarly to variability in the outcome function, extending boxes where the outcome is near-constant and shrinking them where it changes rapidly. For experiments conducted entirely with discrete data, MIP AHB and Fast AHB constructed identical boxes. Lastly, ITE comparisons between the methods show little to no difference in most simulations.

\begin{table*}[!hbtp]
\centering
\caption{Functions used for treatment and confounding in experiments. Continuous covariates are denoted by $x$ and discrete covariates by $w$. There are $p_c$ continuous covariates and $p_d$ discrete covariates.}
\begin{tabular}{c|c|c|c|c|c|c|c}
\hline\hline
& None & Const & Box & Linear & Quad & Binary & Mixed \\ \cline{1-8}
$g(\bx_i)$ or $h(\bx_i)$ & 0 & 1 & $\sum_j \mathbb{I}\{0.5 < x_{ij}\}$ & $\sum_j x_{ij}$ & $\sum_j x_{ij} ^ 2$ & $ w_{ij} $ & $\sum_j (x_{ij} + w_{ij})$\\ 
\cline{1-8}
$(p_c, p_d)$ & (0, 0) & (0, 0) & (2, 0) & (2, 0) & (2, 0) & (0, 1) & (1, 1)\\
\cline{1-8}
\end{tabular}
\label{tab:sim_setup}
\end{table*}

\begin{table*}[!hbtp]
\caption{Mean absolute error as proportion of ATT for estimating ITE of treated units under different confounding regimes. The first column denotes the number of (confounding, treatment, irrelevant) covariates. The second column denotes the confounding and treatment functions, $g$ and $h$ respectively. Either MIP or Fast performs best in almost all simulation types. NA denotes inability to make any matches; bold denotes lowest error attained in that setting.}
\centering
\resizebox{\textwidth}{!}{
\begin{tabular}{c|l|cc|c|c|cccccc} 
\hline\hline
  & & \multicolumn{2}{c|}{AHB} & Black Box & Benchmark & \multicolumn{6}{c}{Matching}\\
  \cline{2-6}
   $p$ & \diagbox{$g$ / $h$}{Method} & MIP & Fast & BART & Best CF & CEM & \makecell{Full\\Matching} & GenMatch & Mahal & \makecell{Nearest\\Neighbor} & Prognostic \\ 
  \hline
  (0, 0, 2) &  None / Const & 0.09 & 0.05 & \textbf{0.04} & 0.25 & 1.01 & 0.32 & 0.36 & 0.34 & 0.37 & 0.25 \\ \hline
  (2, 0, 0) & Box / Const & \textbf{0.11} & 0.16 & 0.24 & 0.24 & 0.24 & 3.03 & 0.66 & 0.62 & 3.05 & 0.29 \\ \hline
  (2, 0, 0)&  Linear / Const & 0.17 & 0.22 & \textbf{0.14} & 0.26 & 0.23 & 0.82 & 0.38 & 0.36 & 0.91 & 0.28 \\ \hline
  (2, 0, 0) & Quad / Const & 0.10 & \textbf{0.04} & 0.08 & 0.25 & 0.22 & 0.42 & 0.38 & 0.37 & 0.45 & 0.27 \\ \hline
  (2, 0, 4) & Quad / Const & \textbf{0.02} & \textbf{0.02} & \textbf{0.02} & 0.16 & NA & 0.21 & 0.12 & 0.11 & 0.24 & 0.04 \\ \hline
  (1, 1, 0) & Box / Box & \textbf{0.30} & 0.45 & 0.65 & 0.73 & 0.58 & 2.59 & 2.37 & 1.02 & 2.30 & 0.94\\ \hline
  (1, 0, 1) & Binary / Const & \textbf{0.02} & \textbf{0.02} & \textbf{0.02} & 0.09 & \textbf{0.02} & 0.12 & 0.49 & 0.10 & 0.10 & 0.09 \\ \hline
  (1, 1, 6) & Binary / Binary & \textbf{0.06} & \textbf{0.06} & 0.09 & 0.17 & 0.20 & 0.71 & 0.97 & 0.27 & 0.61 & 0.18 \\ \hline
  (2, 0, 0) & Mixed / Const & 0.07 & 0.12 & \textbf{0.06} & 0.09 & 0.12 & 0.48 & 0.15 & 0.15 & 0.55 & 0.10 \\ \hline
\end{tabular}}
\label{tab:exp_results}

\end{table*}

\section{APPLICATION}\label{sec:application}
We apply our methodology to replicating a study of the effect of work training programs on future earnings originally conducted by \parencite{lalonde1986,dehejia1999, dehejia2002}.
This dataset includes an experimental sample (from the 1975-76 National Supported Work (NSW) program where treatment units received a work training program), and two observational samples (constructed by combining samples from the Panel Study of Income Dynamics (PSID) and from the Current Population Survey (CPS)). Further details about the datasets are in the Supplement. Matching methods can be evaluated on how well they can reconstruct the unbiased ATT estimate from the experimental sample, by matching treated units from the experiment to control units from the observational samples. Matching covariates include income before the training program, race, years of schooling, marital status, and age. We focus on the task of estimating the in-sample ATT, and therefore match each treated unit $i$ to at least one control unit from each dataset, and no other treated unit. The resulting ITE estimates are then averaged to compute an ATT estimate. We employ MIP AHB, as the data is small enough to do so. Since we do not match any other treated units to each unit $i$, we set $\gamma_1 = 0$, and focus on finding control matches.

\begin{table}[ht]
\centering
\caption{US \$ estimates of the effect of a training program on future earnings from two observational (CPS, PSID) control samples. Methods estimate the ATT by matching treated experimental units to observational control units. The unbiased experimental ATT from the NSW data estimate is \$1794. Estimates closer to this value are better. Estimation error in parentheses.}
\label{tab:lalonde}
\centering
\begin{tabular}{l|ll}
\hline\hline
 \diagbox{Method}{Dataset} & CPS & PSID  \\ 
 \hline
 Adaptive Hyper-box & \textbf{1720 (-75)} & \textbf{1762 (-32)} \\ 
 \hline
 Naive & -7729 (-9523) & -14797 (-16591) \\ 
 \hline
 Full Matching & 708 (-1087) & 816 (-978) \\ 
 \hline
 Prognostic & 1319 (-475) & 2224 (429) \\ 
 \hline
 CEM & 3744 (1950) & -2293 (-4087) \\ 
 \hline
 Mahalanobis & 1181 (-614) & -804 (-2598) \\ 
 \hline
 Nearest Neighbor & 1576 (-219) & 2144 (350) \\ 
   \hline
\end{tabular}

\end{table}

\begin{figure}[h]
    \centering
    \includegraphics[width=.7\columnwidth]{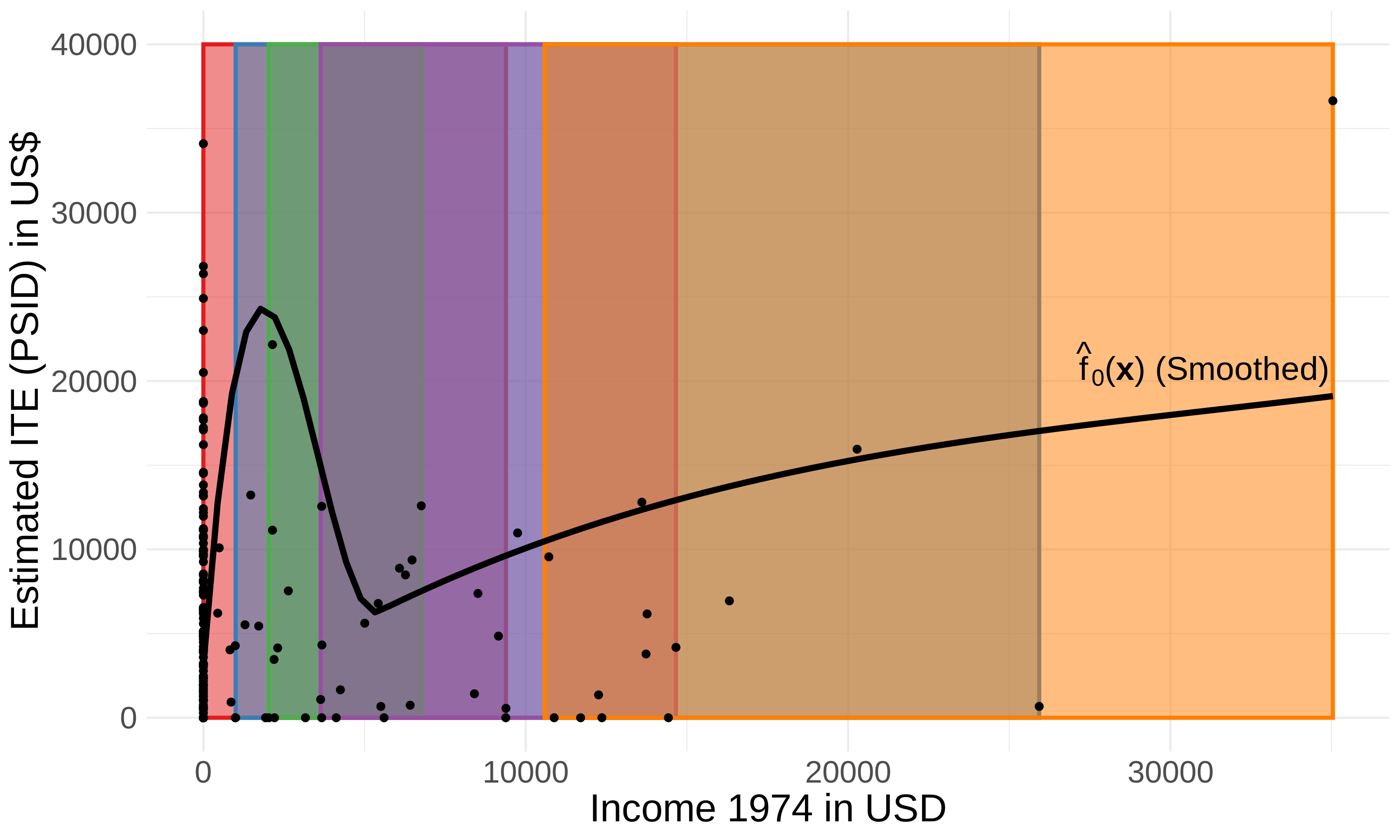}
    \caption{Relationship between pre-treatment income and estimated ITE. The solid black line is a smoothed estimate of the control response as a function of pre-treatment income. The colored boxes are five sample boxes created by Adaptive Hyper-Boxes.}
    \label{fig:lalondeboxes}
\end{figure}

We compare Adaptive Hyper-Boxes to other matching methods estimating the ATT from the observational samples, shown in Table \ref{tab:lalonde}. \textit{The ATT estimates that AHB produces using both observational datasets are comparable to the estimate from the experimental sample.} Most other methods fail to produce estimates of the same quality as AHB on either dataset. Figure \ref{fig:lalondeboxes} displays sample boxes constructed by MIP AHB on one of the matching covariates, together with a smoothed version of the estimated ITE and predicted outcome. Our method behaves as expected, making many small and close boxes where the predicted outcome function grows rapidly, and wider boxes where it does not. 

Table \ref{tab:education} in the supplement also presents treatment effect estimates at different values of pre-treatment years of schooling. We see that years of schooling does indeed moderate the treatment effect, as individuals with fewer years of schooling are estimated to either benefit less than individuals with more years of schooling, or even lose income, after the work training program. Lastly, Table \ref{tab:mgs} shows sample matched groups produced by AHB. 

\section{DISCUSSION}\label{sec:discussion}
Adaptive Hyper-Boxes Matching is a useful alternative to other matching methods. It learns matched groups adaptively, works for mixed categorical and continuous datasets, and produces low-variance matched groups that can be described with interpretable rules. Code implementing AHB is available at \texttt{github.com/almost-matching-exactly/\\Adaptive-Binning}. Hyper-boxes have a long history of successful usage in regression and classification problems. They can produce interpretable predictions which we have now leveraged to produce interpretable matches in the context of causal inference.

\subsubsection*{Acknowledgements}
This work was supported in part by NIH award R01EB025021, NSF awards IIS-1552538 and IIS-1703431, a DARPA award under the L2M program, and a Duke University Energy Initiative ERSF grant.

\clearpage
\printbibliography[title= References]


\clearpage
\normalsize
\section{Supplement}
\subsection{Algorithm Details}\label{sec:algdetails}
\paragraph{Full MIP Formulation} We provide the fully linear form of MIP AHB that can be solved by any MIP solver. We give the MIP formulation to match one unit.
\begin{align*}
    \min\limits_{\bH_i}&\biggl\{
    \gamma_1\sum_{k=1}^nw_{ik}\left|\fhat_1(\bxts_i) - \fhat_1(\bxts_k)\right|\nonumber
    \\&+\gamma_0\sum_{k=1}^nw_{ik}\left|\fhat_0(\bxts_i) - \fhat_0(\bxts_k)\right|\nonumber-\beta\sum_{k=1}^n w_{ik}\label{Eq:MIPloss}\biggr\}\\
    &\mbox{Subject to: }\\
    & \forall\, j,k\;: a_{ij}, b_{ij} \in \mathbb{R}, w_{ik}, u_{jk}, v_{kj} \in \{0,1\}\;\\
    &\forall\, j\;:a_{ij} \leq x_{ij}\\
    &\forall\, j\;:-b_{ij} \leq -x_{ij}\\
    &\forall\, j,k\;:Mu_{kj} + a_{ij} \leq M + x_{kj}\\
    &\forall\, j,k\;:-Mu_{kj} - a_{ij} \leq - x_{kj}\\
    &\forall\, j,k\;:Mv_{kj} - b_{ij} \leq M - x_{kj}\\
    &\forall\, j,k\;:-Mv_{kj} + b_{ij} \leq x_{kj}\\
    &\forall\, k\;: \sum_{j=1}^p u_{ij} + \sum_{j=1}^p v_{ij} - Mw_{ik} \leq 2p - 1\\
    &\forall\, k\;: -\sum_{j=1}^p u_{ij} - \sum_{j=1}^p v_{ij} + Mw_{ik} \leq -2p + M\\
    &\qquad \sum_{k=1}^n w_{ik} \geq m.
\end{align*}
Here $a_{ij}$ and $b_{ij}$ are decision variables respectively representing the lower and upper bounds of the hyper-boxes being constructed, for each covariate. The additional decision variables are needed to implement logical constraints of the form in  \eqref{Eq:MIPC2}. In the above, $M$ is a large positive constant. 

\paragraph{Pre-processing for increased computation speed} There are two ways of preprocessing data before use with MIP AHB to increase computation speed. All the following is for the problem of matching one treated unit $i$ to possible units $k=1,\dots,n$. \\
1. Since it is likely that only units close to $i$ in terms of the MIP objective: $L_{ij} = |\fhat_1(\bx_i) - \fhat_1(\bx_k)| + |\fhat_0(\bx_i) - \fhat_0(\bx_k)|$ will be matched together, it is possible to pre-process data to consider only units that have a value of $L_{ik}$ that is sufficiently small. This can be achieved either by pre-computing $L_{ik}$ for all $k$ and then by excluding units with $L_{ik} > \epsilon$, for some pre-defined threshold, $\epsilon$, or by sorting the $k$ candidate units increasingly in $L_{ik}$ and using only the first $d$ as inputs to MIP AHB.\\
2. It is also likely that only units that are close enough to $i$ in terms of $|x_{ij} - x_{kj}|$ in every covariate, $j$, will  be matched to $i$ by MIP AHB, as choosing units that are too far away might introduce too much variance in the resulting box. Because of this, another useful pre-processing step can be to compute the absolute distance of all $k$ units from $i$ in every  covariate, and then choose, as candidates for matching, only units $k$ such that $|x_{ij} - x_{kj}| \leq \epsilon$ for all $j$ and some threshold, $\epsilon$.

\begin{table}[htbp]
\centering
\caption{Error and run time for MIP AHB with different preprocessing methods, using different data generation processes (DGP's). Error is mean absolute error, and run time is mean time in seconds to match all test units. In all simulations there were 300 test units to match, roughly evenly split between treated and control. For each DGP, 5 simulations were run. For sorting preprocessing, the closest 50 treated and 50 control units were chosen. For threshold on $L_{ik}$ the threshold was 0.4, and for threshold on $|x_{ij} - x_{kj}|$ the threshold was 0.5. }
\label{tab:preprocessing}
\begin{tabular}{rr|llll}
  \hline\hline
  \multicolumn{2}{c|}{\diagbox{DGP}{Preprocessing}} & None & Sort on $L_{ik}$ & Threshold $L_{ik}$ & \makecell{Threshold\\ $|x_{ij} - x_{kj}|$} \\ 
  \hline
  Linear & Error & 0.063 & 0.070 & 0.066 & 0.063 \\ 
         & Time & 112.7 &  26.2 &  51.2 &  76.1 \\
         \hline
  Quadratic & Error & 0.065 & 0.065 & 0.064 & 0.066 \\ 
            & Time & 100.3 &  12.5 &  26.1 &  48.9 \\ 
            \hline
  Box & Error & 0.030 & 0.051 & 0.061 & 0.033 \\ 
      & Time & 66.6 &  4.5 &  6.7 & 36.5 \\ 
   \hline
\end{tabular}
\end{table}

Table \ref{tab:preprocessing} shows results for ITE estimation with MIP AHB under some of the data generation models employed in Section \ref{sec:experiments}, and after preprocessing in different ways. The table shows that preprocessing leads to substantial gains in terms of computation time, without almost any decrease in performance in terms of absolute estimation error. 

\paragraph{Fast AHB Pseudocode}
Fast AHB is shown in Algorithm \ref{greedy_algo}. At every iteration, we consider expanding $i$'s box along each covariate (line 2) to the closest value outside the box that is also realized by that covariate in the data (lines 3 -- 4). Ties are broken arbitrarily. For each covariate, if this new value is \emph{above} the upper endpoint of the covariate's current box (line 5), we propose raising the box's upper endpoint (line 6); otherwise, we propose lowering the box's lower endpoint (line 7). To determine whether the new, proposed box $\mathbf{P}$ is good, we see how much the outcome function changes in the \emph{additional} region we would be including in the box if we expanded to $\mathbf{P}$: $\mathbf{P}\backslash \bH_i$ (where $\backslash$ denotes set difference). If these predicted outcomes are similar to one another -- they have low variability -- then the outcome function is relatively constant in the new region and we do not expect to incur much bias from including units that might lie inside. Therefore, we look at how much $\hat{f}$ varies on a grid in this additional region (line 8) and choose to expand along the covariate that yields the lowest variation (lines 9 -- 14). We update $i$'s box and matched group (lines 15 -- 16) and continue expanding in this manner until the variability in $\hat{f}$ jumps above some threshold from the previous iteration (line 1). We also enforce the constraint that each unit is matched to at least one another with opposite treatment (line 1).
\begin{algorithm}[hbtp]
 \textbf{Inputs}: $D^{ts}, \hat{f}_0, \hat{f}_1, c$\\
 \KwResult{A $p$-dimensional hyper-box for unit $i$}
 Initialize $a_{ij} := b_{ij} := x_{ij}$ for all $j$\\
 Initialize $\bH_i^{(0)} = H(\ba_i, \bb_i)$\\
 Initialize $v_*^{(0)} := v_*^{(1)} >> 1$\\
 Initialize $s := 1$\\
 (While the stopping conditions are not met:)\\
\nl \While{$v_*^{(s)} < cv_*^{(s-1)}\textrm{ or }\mathtt{MMG}^{(s-1)}(\bH_i) \cap Y_{i|t_i = 0} = \emptyset$}{
    Initialize $tmp$\;
\nl \For{$j = 1:p$}{
        \nl(Find closest unit $i^*$:)\\
        $i^*_{down} \in \text{argmin}_{i': x_{i'j} < a_{ij}} \min(a_{ij} - x_{i'j})$\;
        $i^*_{up} \in \text{argmin}_{i': x_{i'j} > b_{ij}} \min(x_{i'j} - b_{ij})$\;
        \eIf{$|a_{ij} - x_{i^*_{down}j}| < |x_{i^*_{up}j} - b_{ij}|$}
        {$i^* := i^*_{down}$}
        {$i^* := i^*_{up}$}
        \nl $tmp[j] := i^*$\;
        \nl \eIf{$x_{i^*j} > b_{ij}$}{
        \textrm{(propose expanding box upward)}\\
        \nl $\mathbf{P} := H(\ba_i, (b_{i1}, \dots, x_{i^*j}, \dots, b_{ip}))$\;
        }{
        \textrm{(propose expanding box downward)}\\
        \nl $\mathbf{P} := H((a_{i1}, \dots, x_{i^*j}, \dots, a_{ip}), \bb_i)$\;
        }
        \nl $v_j := \text{var}\{\hat{f}_0(p_k)\} + \text{var}\{\hat{f}_1(p_k)\}$ for a grid of evenly spaced points $p_k \in \mathbf{P} \backslash \bH_i^{(s - 1)}$\;
  }
  \nl $j^* \in \text{argmin}_j v_j$\;
  \nl $i^* := tmp[j^*]$\textrm{ (retrieve point to expand to)}\;
  \nl $v_*^{(s)} := \min v_j$\;
  \nl \eIf{$x_{i^*j} > b_{ij}$}{
    \nl $b_{ij^*} := x_{i^*j^*}$\textrm{ (expand box upward)}\;
    }{
    \nl $a_{ij^*} := x_{i^*j^*}$\textrm{ (expand box downward)}\;
    }
    \nl $\bH_i^{(s)} := H(\ba_i, \bb_i)$\;
  \nl $\mathtt{MMG}^{(s)}(\bH_i):= \{k: \bx_k \in \bH_i^{(s)}\}$\;
  $s = s + 1$\;
 }
 \caption{Fast Approximation to AHB Matching
 \label{greedy_algo}}
\end{algorithm}

\subsection{More Examples of Adaptive Hyper-Boxes}

\begin{figure}[!htbp]
    \centering
    \includegraphics[width=\columnwidth]{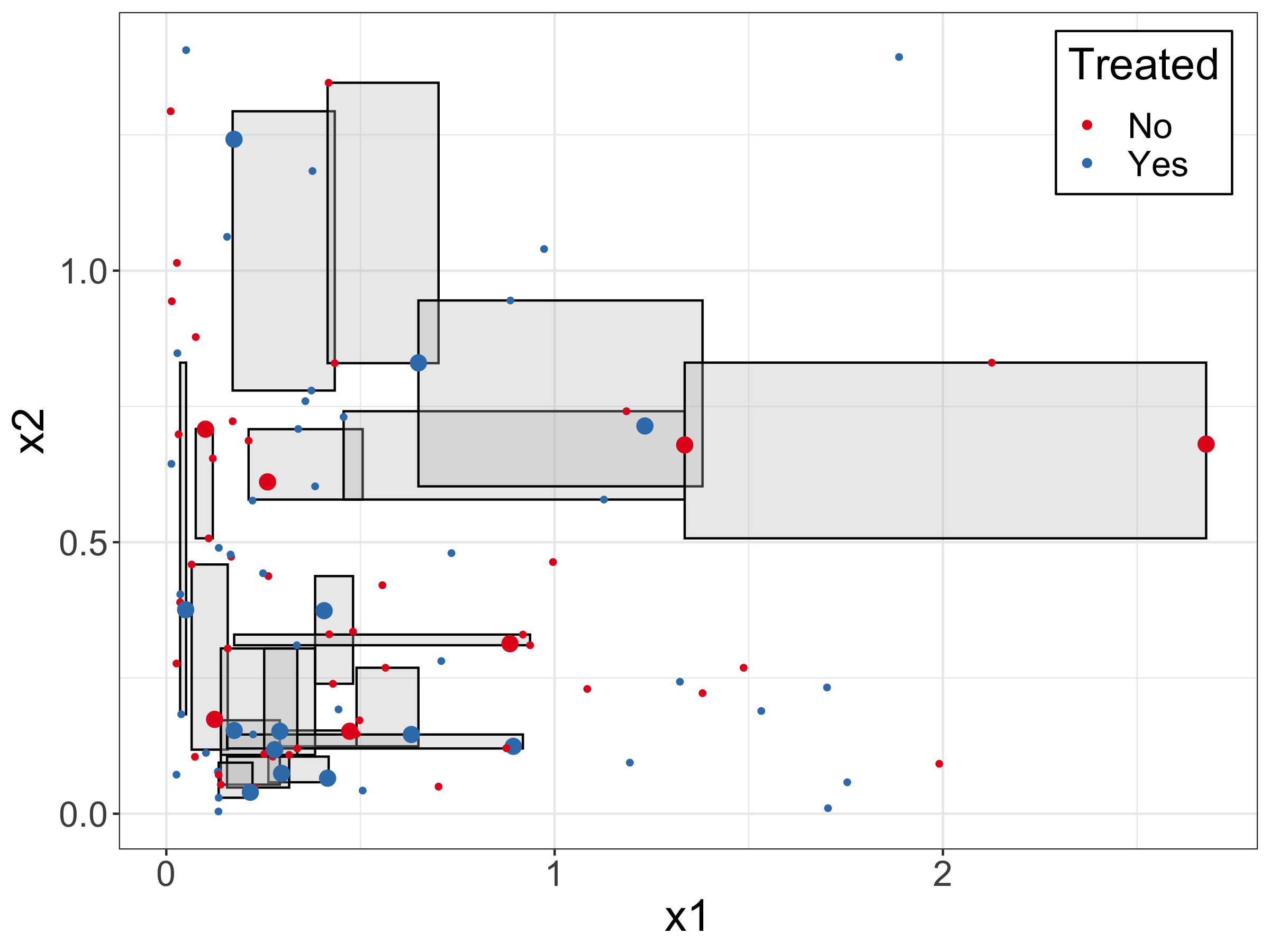}
    \caption{20 sample hyper-boxes made on 100 test units with a logarithmic outcome function.}
    \label{fig:expboxes}
\end{figure}

Figure \ref{fig:expboxes} shows 20 sample hyperboxes made on a simulated dataset of 100 units. In this case the data generating process was:
\begin{align*}
x_{i1}, x_{12} &\sim \operatorname{Exponential}(2)\\
Z_i &\sim \operatorname{Bernoulli}(0.5)\\
\epsilon_i &\sim \operatorname{Normal}(0,1)\\
Y_i &= 3Z_i + \log(x_{i1} + x_{i2}) + \epsilon_i.
\end{align*}
In this case the data is concentrated in a region (bottom left of the figure)  in which the outcome function increases rapidly as a function of the covariates, and therefore boxes are smaller. Towards the top right portion of the figure, the outcome function grows less rapidly with $x_1$ and $x_2$, and therefore larger boxes can be made without losing too much accuracy. This pattern is present in the figure.

\begin{figure}[!htbp]
    \centering
    \includegraphics[width=\columnwidth]{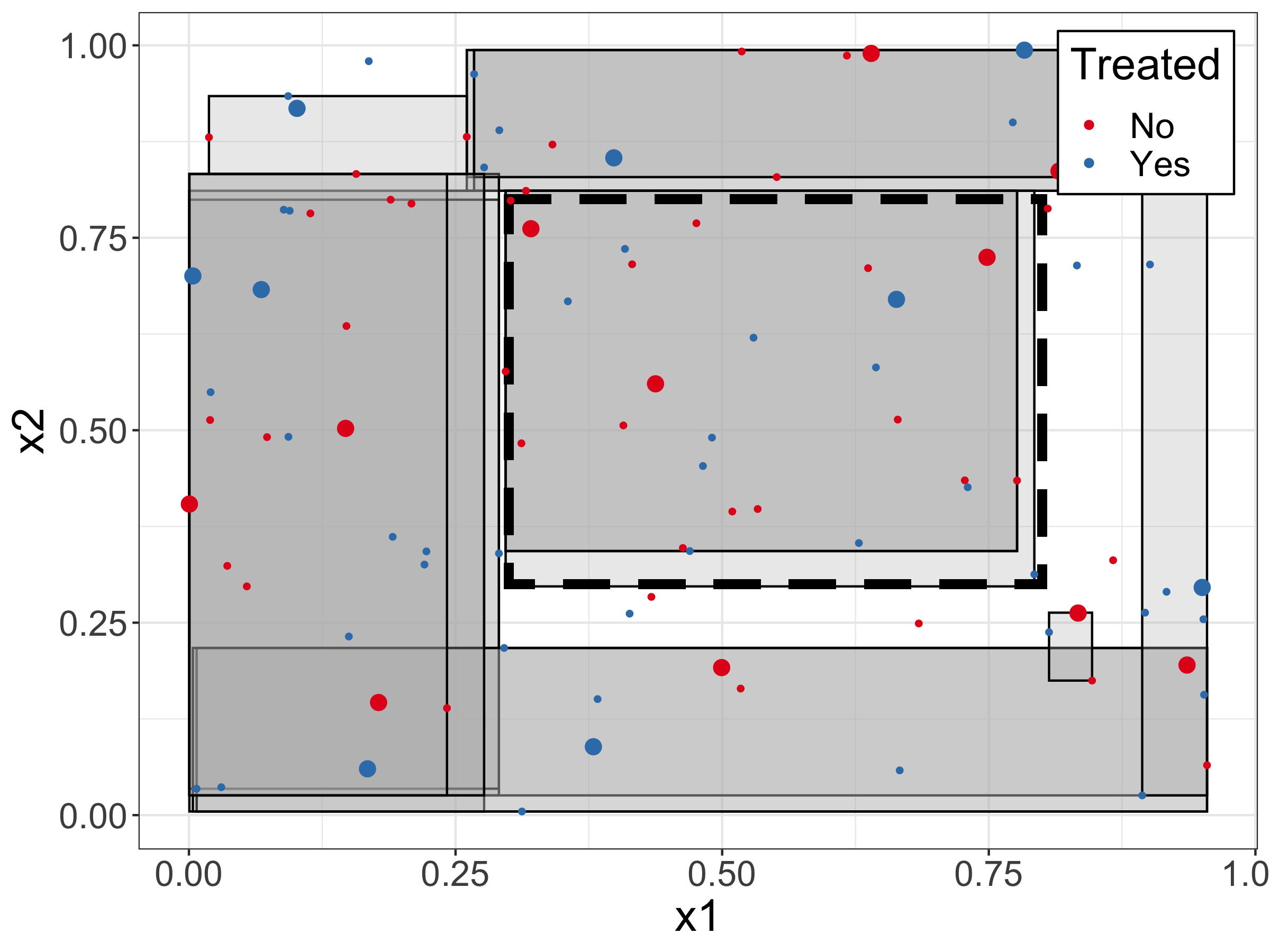}
    \caption{20 sample hyper-boxes made on 100 test units, with a rectangle-shaped outcome function. Units inside of the dashed rectangle experience confounding, and units outside do not. }
    \label{fig:boxboxes}
\end{figure}

Figure \ref{fig:boxboxes} shows 20 sample hyperboxes made on a simulated dataset of 100 units. In this case the data generating process was:
\begin{align*}
x_{i1}, x_{12} &\sim \operatorname{Uniform}(0, 1)\\
Z_i &\sim \operatorname{Bernoulli}(0.5)\\
\epsilon_i &\sim \operatorname{Normal}(0,1)\\
Y_i &= 3Z_i + \ind_{[0.3 < x_{i1}, x_{i2} < 0.8]} + \epsilon_i.
\end{align*}
In this case the treatment effect is constant and only units inside of the square experience confounding, while the units outside do not. Confounding inside the box is constant, so all units inside should be matched together, and all units outside should also be matched together. We see that AHB almsot perfectly replicates this pattern with the boxes it constructs. 

\subsection{Parameter Tuning}

\textbf{MIP AHB}: There are three hyperparameters in our MIP formulation: $\gamma_0, \gamma_1, \beta$. The first two control the weight placed on the outcome function portion of the loss, while the third controls the weight placed on the number of units in the box. These three components are not on the same scale, as the first two components of the loss are on the same scale as the outcome, and the third is a 0/1 binary. Because of this, it is useful to rescale the first two components of the loss to be between 0 and 1 by dividing $\gamma_0$ and $\gamma_1$ by the sample variance of $|\fhat_0(\bx_k)|$ and $|\fhat_1(\bx_k)|$, $k=1,\dots, n$ respectively. This ensures that loss values are standardized and roughly on a 0/1 scale. This aids with comparability and interpretability of the hyperparameters, as a value of standardized $\gamma_0=2$ and $\beta=1$ will mean that the first component of the loss should be weighted roughtly twice as much as the third. 

In some cases, one may choose to avoid giving weight to the treatment outcomes $f_t(\bx_k)$: when an estimator of the form $\tauihat = Y_i(1) - \yhati$ is used, then users might want to prioritize matching to impute good control counterfactual outcomes. The parameters $\gamma_0$ and $\gamma_1$ allow for this differential weighting of objective values.

\paragraph{Hyperparameter tuning} Hyperparameters for both algorithms can be tuned by cross-validation on a separate validation set. Let $D^{vl} = \{(\bxvl_i, \Yvl_i, \Tvl_i)\}_{i=1}^n$ denote this separate validation set. We propose the following hyper-parameter tuning procedure, which chooses the hyper-parameter that minimizes validation loss:
\begin{enumerate}
    \item Construct a set of candidate hyperparameter values $\blambda = (\lambda_1,\dots, \lambda_q).$
    \item For each element of $\blambda$, $\lambda_s$:
    \begin{enumerate}
        \item Use either algorithm to estimate the \textbf{observed outcome} of each validation unit via matching, $i$, that is, output $\yhati$ for control validation units, and $\yhatione$ for all treated validation units. 
        \item Compute validation loss: 
        \begin{align*}
            L^{vl}(\lambda_s) = \sum_{i=1}^n &(\Yvl_i - \yhatione)^2T_i\\
            &+ (\Yvl_i - \yhati)^2(1-T_i).
        \end{align*}
    \end{enumerate}
    \item Choose $\lambda^* \in \text{argmin}_{\lambda^s \in \blambda}L^{vl}(\lambda_s)$.
\end{enumerate}


\subsection{Algorithm Parallelization and Scalability}
As discussed earlier, both MIP AHB and Fast AHB are embarrassingly parallelizable, as they construct the box of any given unit independently of those of others. 

As far as its complexity, for one unit, MIP AHB is a discrete optimization problem with $O(np)$ integer decision variables plus $O(p)$ real decision variables, and $O(np)$ constraints. There are two routes to improve computation speed for MIP AHB. First, computation over the $n$ units can be easily parallelized, as explained above. Second, it is possible to pre-process the data by choosing only units $k$ for which either distance in terms of black-box predictions or raw covariate values between $i$ and $k$ is below a pre-defined threshold: this will leave $d \leq m$ candidate units for matching to unit $i$, and will bring the number of integer decision variables and constraints in the MIP down to $O(dp)$. This step is discussed previously in more detail previousy in this supplement, showing it can be used to decrease the MIP runtime for large datasets with negligible loss in performance.

Fast AHB provides a computational advantage over MIP AHB. Its slowest element comes from repeated evaluation of $\hat{f}_0$ and $\hat{f}_1$; when constructing a box for a single unit, we must, at every iteration, generate $p$ estimates of the variability of the outcome function in the region we are expanding into (see line 8 in Algorithm \ref{greedy_algo}). Generating these predictions using $\hat{f}_0$ and $\hat{f}_1$ to do so is the computational bottleneck of the method, depending heavily on the speed of the predictive method employed. 

\begin{figure}
    \centering
    \includegraphics[height=\columnwidth]{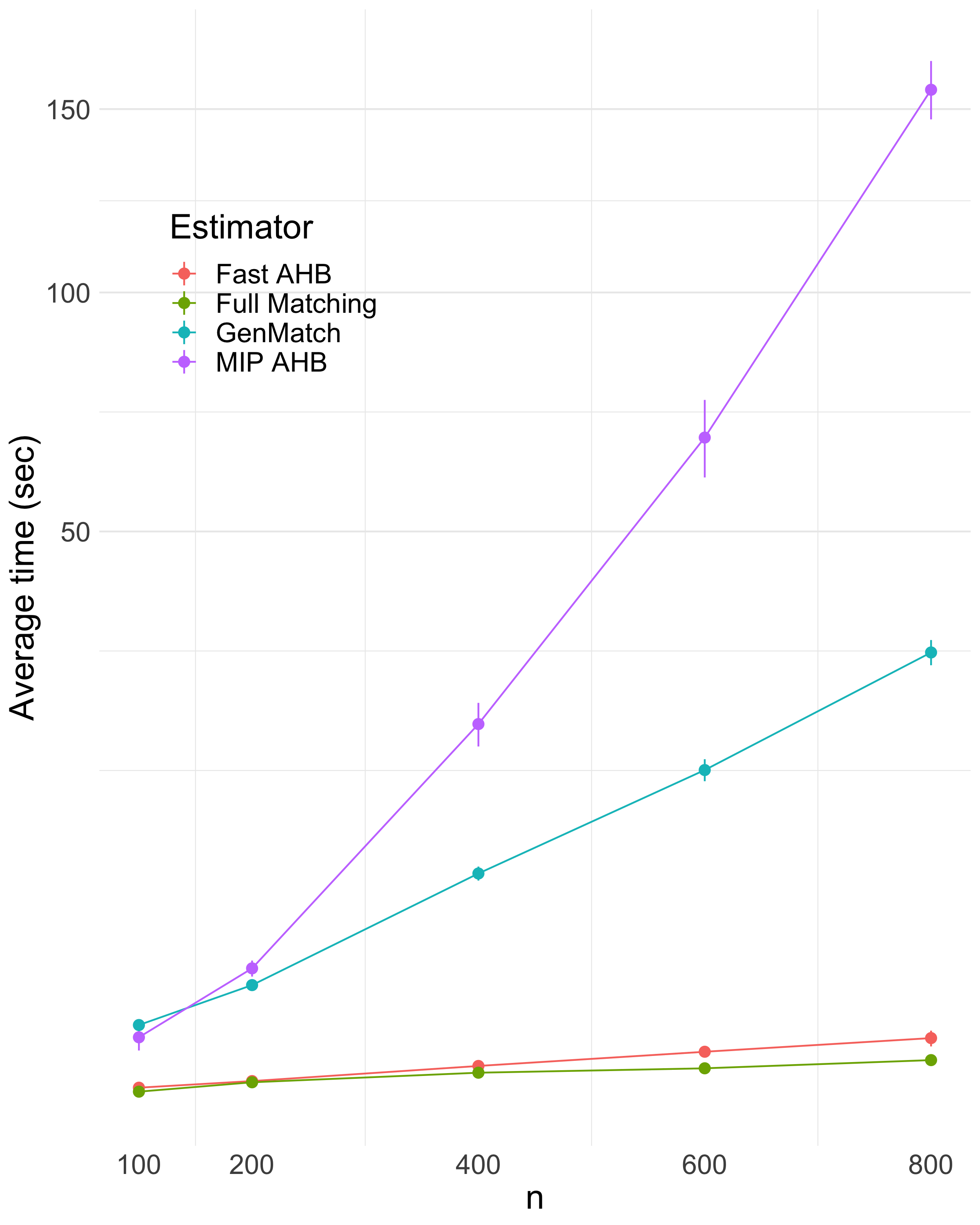}
    \caption{Average runtimes of various methods across 10 simulations for fixed $p = 2$ and increasing $n$. Error bars denote standard deviations of runtimes across the simulations.
    \label{fig:scale_n_discrete}}
\end{figure}

\begin{figure}
    \centering
    \includegraphics[height=\columnwidth]{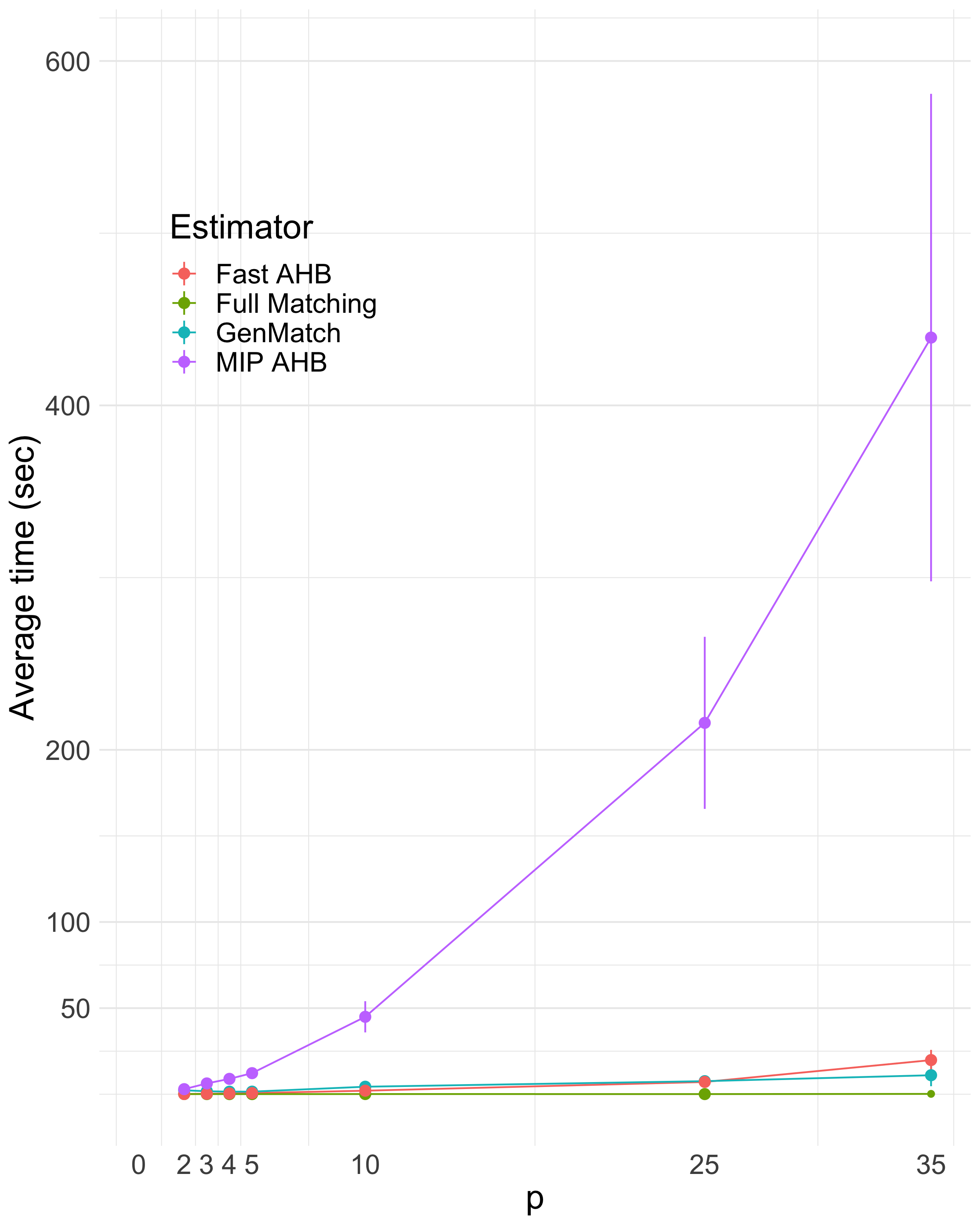}
    \caption{Average runtimes of various methods across 10 simulations for fixed $n = 200$ and increasing $p$. Error bars denote standard deviations of runtimes across the simulations.}
    \label{fig:scale_p_discrete}
\end{figure}

Despite the additional computation required from either MIP AHB or Fast AHB tends to perform relatively quickly, finishing in under 10 minutes for hundreds of observations and tens of covariates when run in serial on a laptop (see Figures \ref{fig:scale_n_discrete} and \ref{fig:scale_p_discrete}). (The actual time in parallel would divide this 10 minutes by $n$, the number of units in the full dataset.)

While MIP AHB performs with similar speed for either discrete or continuous variables,
Fast AHB tends to perform faster on discrete covariates, because for each $k$-level discrete covariate, it can take a maximum of $k - 1$ steps, versus $n - 1$ for continuous covariates. For this reason, it also tends to scale better in $n$ than in $p$. 


\subsection{Additional Experiment Details}
\paragraph{General Details}
For both MIP AHB and Fast AHB, we employ Bayesian Additive Regression Trees (BART), as implemented in the \pkg{dbarts} package in \proglang{R}, to learn $\hat{f}_0$ and $\hat{f}_1$ and supply predictions. We perform cross-validation on the training set in order to determine the number of trees and the variance parameter for the prior distribution on the leaves. 

Genetic Matching was performed via the \pkg{Matching} package; Mahalanobis Distance matching, Propensity Score matching, and Full Matching were performed via the \pkg{MatchIt} package; and CEM was performed using the \pkg{cem} package, all of which are \proglang{R} packages. We implemented prognostic score matching ourselves, using the predictions attained from BART. 

All results are reported as averages across 10 simulations with $n = 600$ units, 400 of which were used to train BART for Prognostic, MIP AHB, and Fast AHB; other methods, none of which use any outcome information, were permitted to train and make matches using all units, though only ITEs for test units are reported. 

\subsection{Additional Experiment Results}
\paragraph{CEM, AHB, and Exact Matching}
In the body, we observed that running AHB on data with two covariates, only one of which was relevant to the outcome, resulted in exact-matching of units on the \emph{relevant} covariate alone. CEM, on the other hand, creates exact matches on all covariates. For two covariates, this has no practical impact. But for greater numbers of covariates, not only will CEM struggle to make matches, while AHB will not, but the variance of its estimates will increase as it creates more partitions of the space than necessary (along irrelevant covariates) to estimate the treatment effect without incurring bias. Table \ref{tab:cem_ahb} shows mean absolute error of CEM and the AHB methods, along with the proportion of data for which they fail to create matches, as the dimensionality of the space increases. These results are averaged across 5 simulations each. 

\begin{table}[]
\caption{Comparison of CEM and AHB performance on binary variables. As number of variables increases, CEM is increasingly unable to match units as highlighted in bold in the lower part of Row 1 of the table, and yields progressively worse treatment effect estimates, given that it tries to match exactly even on irrelevant covariates. AHB methods do not suffer from this problem.}
\resizebox{\columnwidth}{!}{
\begin{tabular}{lc|c|c|c|c|c}
\hline\hline
&  & \multicolumn{5}{c|}{Number of Covariates}\\ \hline
\textbf{Estimator} & & 2 & 3 & 5 & 8 & 10 \\ \hline
CEM & \makecell{Error\\\% Missing}     & \makecell{0.02\\0} & \makecell{0.02\\0} & \makecell{0.04\\0} & \makecell{0.09\\\textbf{32}} & \makecell{0.13\\\textbf{77}} \\ \hline
MIP AHB & \makecell{Error\\\% Missing} & \makecell{0.02\\0} & \makecell{0.01\\0} & \makecell{0.02\\0} & \makecell{0.02\\0} &  \makecell{0.02\\0} \\ \hline
Fast AHB & \makecell{Error\\\% Missing} & \makecell{0.02\\0} & \makecell{0.01\\0} & \makecell{0.02\\0} & \makecell{0.02\\0} &  \makecell{0.02\\0} \\ \hline
\end{tabular}}
\label{tab:cem_ahb}
\end{table}

\paragraph{Correlated Covariates}
We study the performance of AHB when the covariates are highly correlated. We simulate the $\bX_i$ independently from a Gaussian latent factor model with two latent factors, yielding an average pairwise correlation between covariates of about 0.7, across simulations. Other aspects of the simulation are as in Section 3. Simulations results are presented in Table \ref{tab:corr_covs}. AHB methods tend to outperform all methods but the black box, BART. This is unsurprising as AHB leverages the predictive power of the underlying black box to create accurate and interpretable treatment effect estimates. If the black box performs well in settings with high correlation, so will AHB. 

\paragraph{High Dimensional Covariates}
We study the performance of AHB on higher dimensional data than that considered in Section 3. The simulation setup is the same and we consider two Linear/Linear settings with twenty covariates. In the first, only two contribute to the outcome and treatment effect; in the second, all twenty are relevant. Results are presented in Table \ref{tab:high_dim}.

As with correlated covariates, discussed in the previous paragraph, as long as the black box is able to perform well in high dimensional settings, so will AHB. We see that BART has no trouble learning linear functions of 20 covariates or identifying irrelevant covariates. The accurate estimates produced by BART therefore allow MIP AHB and Fast AHB to appropriately coarsen the space, making larger boxes in regions of little outcome variation and vice versa.

\begin{table*}[!hbtp]
\caption{Mean absolute error as proportion of ATT for estimating ITE of treated units under different confounding regimes. The first column denotes the number of (confounding, treatment, irrelevant) covariates. The second column denotes the confounding and treatment functions, $g$ and $h$ respectively. AHB methods outperform all but the black box BART in all simulation types. Bold denotes lowest error attained in that setting.}
\centering
\resizebox{\textwidth}{!}{
\begin{tabular}{c|l|cc|c|c|cccccc} 
\hline\hline
  & & \multicolumn{2}{c|}{AHB} & Black Box & Benchmark & \multicolumn{6}{c}{Matching}\\
  \cline{2-6}
   $p$ & \diagbox{$g$ / $h$}{Method} & MIP & Fast & BART & Best CF & CEM & \makecell{Full\\Matching} & GenMatch & Mahal & \makecell{Nearest\\Neighbor} & Prognostic \\ 
  \hline
  (4, 4, 0) &  Linear / Linear & 0.80 & 1.12 & \textbf{0.18} & 1.12 & 0.94 & 4.33 & 4.64 & 1.51 & 3.81 & 1.12 \\ \hline
  (4, 4, 0) & Box / Box & 0.12 & 0.16 & \textbf{0.07} & 0.12 & 0.22 & 0.34 & 0.16 & 0.14 & 0.15 & 0.13 \\ \hline
  (4, 4, 0)&  Quad / Quad & 0.09 & 0.09 & \textbf{0.06} & 0.12 & 0.22 & 0.34 & 0.13 & 0.14 & 0.15 & 0.13 \\ \hline
\end{tabular}}
\label{tab:corr_covs}

\end{table*}

\begin{table*}[!hbtp]
\caption{Mean absolute error as proportion of ATT for estimating ITE of treated units under different confounding regimes. The first column denotes the number of (confounding, treatment, irrelevant) covariates. The second column denotes the confounding and treatment functions, $g$ and $h$ respectively. AHB methods outperform all but the black box BART in all simulation types. Bold denotes lowest error attained in that setting.}
\centering
\resizebox{\textwidth}{!}{
\begin{tabular}{c|l|cc|c|c|cccccc} 
\hline\hline
  & & \multicolumn{2}{c|}{AHB} & Black Box & Benchmark & \multicolumn{6}{c}{Matching}\\
  \cline{2-6}
   $p$ & \diagbox{$g$ / $h$}{Method} & MIP & Fast & BART & Best CF & CEM & \makecell{Full\\Matching} & GenMatch & Mahal & \makecell{Nearest\\Neighbor} & Prognostic \\ 
  \hline
  (20, 20, 18) &  Linear / Linear & 0.09 & 0.12 & \textbf{0.08} & 0.08 & NA & 0.44 & 0.36 & 0.22 & 0.38 & 0.10 \\ \hline
  (10, 10, 0) & Linear / Linear & 0.15 & 0.20 & \textbf{0.07} & 0.17 & NA & 0.29 & 0.29 & 0.22 & 0.26 & 0.19 \\ \hline
\end{tabular}}
\label{tab:high_dim}

\end{table*}

\subsection{Additional Application Details and Results}


The experimental data is the adapted sample of the National Supported Work Program (NSW) that ran in 1975-76 from \cite{dehejia1999}. There are 185 treated units and 260 control units in this dataset, the former were assigned to receive a work training program uniformly at random. The other two datasets employed are 3 samples from the Panel Study of Income Dynamics (PSID) and 3 from the Current Population Survey (CPS), which we combine together into two full-sample datasets as done by \cite{dehejia2002}. These two datasets contain large pools of control units ($n CPS = 15992$, $n PSID = 2490$) that experimental treated units can potentially be matched to. Matching covariates include income before the training program, race, years of schooling, martial status, and age. We focus on the task of estimating the in-sample ATT, and therefore match each treated unit $i$ to at least one control unit from each dataset, and no other treated unit. We employ the MIP version of our method, as the data is small enough to allow it without parallelization on a laptop. Since we do not match any other treated units to each unit $i$, we set $\gamma_1 = 0$, and focus on finding control matches. 

Additional results for our application section are presented in Tables \ref{tab:education} and \ref{tab:mgs}.
\begin{table}[t]
\centering
\caption{Conditional Average Treatment Effect estimates for years of schooling. Columns 2 and 3 include estimates based on the PSID and CPS control populations in US \$. Modal lower bounds (Lb) and upper bounds (Ub) are the modal bounds for a years of schooling across all matched units. That is, in Row 1, a plurality of treated individuals with 4 years of schooling were matched to individuals who had between 3 and 5 years of schooling. }
\label{tab:education}
\begin{tabular}{rrrrrrr}
  \hline\hline
   \makecell{Years of \\schooling} & \makecell{CPS CATE\\ estimate} & \makecell{PSID CATE\\ estimate} & Modal Lb & Modal Ub \\ 
  \hline
   4 &  1884 & 3550 & 3 & 5 \\ 
   5 &  6186 & 6882 & 3 & 6 \\ 
   6 &  0 & -1161 & 0 & 8 \\ 
   7 &  -3887 & 16 & 4 & 8 \\ 
   8 &  -2545 & -1831 & 3 & 9 \\ 
   9 &  1636 & 402 & 8 & 12 \\ 
   10 & -462 & 516 & 9 & 11 \\ 
   11 & 2093 & 2927 & 10 & 12 \\ 
   12 & 1671 & 712 & 8 & 13 \\ 
   13 & 5656 & 6941 & 12 & 18 \\ 
   14 & 15471 & 12771 & 7 & 15 \\ 
   15 & 3820 & 6683 & 6 & 18 \\ 
   16 & -1143 & -4506 & 10 & 18 \\ 
   \hline
\end{tabular}

\end{table}

\begin{table}[ht]
\centering
\caption{Two sample matched groups produced by our method using the PSID control sample. All but the last column are matching covariates. Bolded lines are treatment units, other lines represent control units.  }
\label{tab:mgs}
\resizebox{\textwidth}{!}{
\begin{tabular}{rrrrrrrr|r}
  \hline\hline
  Age & Education & Black & Hispanic & Married & No H.S. Degree & Income 1974 & Income 1975 & \makecell{Income 1978\\(outcome)}  \\ 
  \hline
  \multicolumn{9}{c}{Matched group 1}\\
  \hline
  \bf 28 & \bf 9 & \bf 1 & \bf 0 & \bf 0 & \bf 1 & \bf 0 & \bf 0  & \bf 10694 \\ 
  \hline
  47 & 12 & 0 & 0 & 0 & 0 & 0 & 0 & 0 \\ 
  44 & 12 & 0 & 0 & 0 & 0 & 0 & 0 & 0 \\ 
  44 & 12 & 0 & 0 & 0 & 0 & 0 & 0 & 0 \\ 
   48 & 11 & 0 & 0 & 0 & 1 & 0 & 0 & 0 \\ 
   47 & 12 & 0 & 0 & 0 & 0 & 0 & 0 & 0 \\ 
  \hline
\multicolumn{9}{c}{Matched group 2}\\
  \hline
  \bf 33 & \bf 12 & \bf 1 & \bf 0 & \bf 1 & \bf 0 & \bf 20280 & \bf 10941 & \bf 15953 \\ 
  \hline
  28 & 14 & 1 & 0 & 1 & 0 & 17633 & 12532 & 16255 \\ 
  32 & 14 & 1 & 0 & 1 & 0 & 19593 & 12335 & 14777 \\ 
  27 & 16 & 0 & 0 & 1 & 0 & 19593 & 12532 & 17721 \\ 
  38 & 12 & 0 & 0 & 1 & 0 & 19593 & 13965 & 16286 \\ 
  35 & 12 & 1 & 0 & 1 & 0 & 21552 & 14323 & 15639 \\ 
  46 & 10 & 0 & 1 & 1 & 1 & 17633 & 10742 & 5911 \\ 
  28 & 12 & 0 & 0 & 1 & 0 & 17829 & 11100 & 20688 \\ 
  28 & 13 & 0 & 0 & 1 & 0 & 17876 & 15397 & 22453 \\ 
  23 & 12 & 0 & 0 & 1 & 0 & 19397 & 12532 & 25121 \\ 
  44 & 12 & 0 & 0 & 1 & 0 & 20896 & 12174 & 20963 \\ 
   \hline
\end{tabular}}
\end{table}

\subsection{Confidence Intervals for ITE Estimates}
In this section, we investigate the variance associated with our ITE estimates, and resulting confidence intervals for the true effect. The proper way to estimate variance when matching is an open question. One approach is to use treatment effect estimators with known asymptotic distributions whose variances can be estimated \parencite{abadieimbens2006}. Another is to use bootstrap or subsampling -based approaches \parencite{bootstrap_failure, weighted_bootstrap}. Here, our goal is to empirically study the coverage of confidence intervals constructed around our estimates under a variety of variance-estimation approaches. Specifically, we consider six different constructions of confidence intervals, described below. Three are normal approximations that use different estimates of the variance, two rely on resampling the matching data, and one uses posterior samples from BART to construct a credible interval:
\begin{enumerate}
\item \textbf{NA with BART Variance}: Normal approximation where the variance is the posterior mean of the variance from BART. 
\item \textbf{NA with True Variance}: Normal approximation where the variance is the true unit variance.
\item \textbf{NA with Conservative Variance}: Normal approximation where the variance is the max of: 1. twice the variance of control outcomes in the matched group, 2. twice the variance of treated outcomes in the matched group, 3. the sum of the variances of treated and control outcomes in the matched group.
\item \textbf{Bootstrap CI}: The CI obtained by bootstraping the matching data. The matched group is held fixed, and the units within it are resampled with replacement. 
\item \textbf{Subsampling CI}: The CI obtained by subsampling the matching data. The matched group is held fixed, and subsamples of units therein are taken. 
\item \textbf{BART Cred I}: BART posterior credible interval
\end{enumerate}

\begin{figure}[ht]
    \centering
    \includegraphics[width=.9\columnwidth]{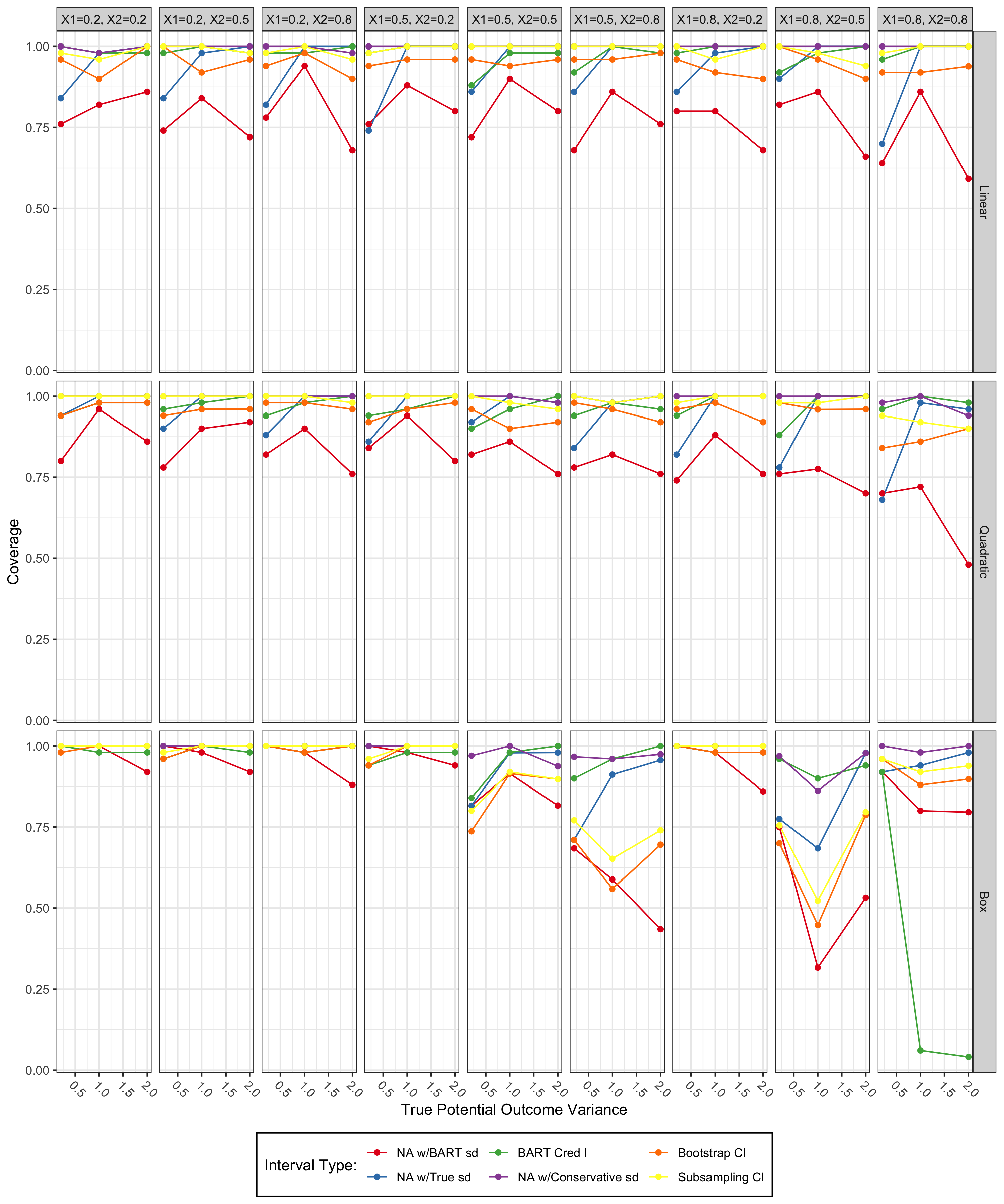}
    \caption{Coverage of 95\% confidence intervals for the true ITE across different variance estimates and simulation settings. The subsampled confidence intervals and those from a normal approximation with a conservative variance estimate nearly always attain 95\% coverage.}
    \label{fig:coverage}
\end{figure}

We consider simulation settings from Section 3, in which $X_1, X_2 \stackrel{ind}{\sim} U(0, 1)$ and confounding and treatment effect modification are both: 1. Linear, 2. Quad, and 3. Box. Within each of these settings, we look at the coverage as a function of both $X_1, X_2$ and the true variance of the potential outcomes (the variance of $\epsilon$ in step 4 of the data generating process described in Section 3). After simulating data in the above manner, we run MIP AHB to generate ITE estimates and then compare the variance and resulting 95\% confidence intervals using the six methods outlined above. Results, presented in Figure \ref{fig:coverage}, are averages across 50 simulations using $n = 600$ units, $400$ of which were used for training.

The normal approximations using the true or BART estimates of variance tend to undercover. This might be expected in the second case given that the BART variance estimate is not coupled with its estimate of the mean. The BART credible interval sidesteps this issue and outperforms the previous estimators by using BART's estimate of variance to expand around its point estimate of the mean. However, it still undercovers in some settings, typically those in which the potential outcome variance is low. The bootstrapped confidence intervals are more consistent, though they slightly undercover. The subsampled and conservative confidence intervals nearly always cover in at least 95\% of the simulations. We therefore recommend either subsampling or using the conservative confidence intervals in order to express uncertainty about ITE estimates. 

\printbibliography[title={Supplement References}]
\end{document}